\documentclass[aps,pra,superscriptaddress,bibnotes]{revtex4-2}

\usepackage[T1]{fontenc}

\usepackage{natbib}
\usepackage{graphicx}
\usepackage{bm}
\usepackage{color}
\usepackage{amsmath}
\usepackage{gensymb} 
\usepackage{physics}
\usepackage{tabularx}
\usepackage{hyperref}

\newcommand{\um}{$\mu$m}
\newcommand{\us}{$\mu$s}
\newcommand{\Na}{{\rm Na}}
\newcommand{\Cs}{{\rm Cs}}

\newcommand{\CUAaddress}{Harvard-MIT Center for Ultracold Atoms, Cambridge, Massachusetts 02138, USA}
\newcommand{\HarvardPhysicsAddress}{Department of Physics, Harvard University, Cambridge, Massachusetts 02138, USA}
\newcommand{\HarvardChemistryaddress}{Department of Chemistry and Chemical Biology, Harvard University, Cambridge, Massachusetts 02138, USA}
\newcommand{\JILAAddress}{JILA, National Institute of Standards and Technology and Department of Physics, University of Colorado,
Boulder, Colorado 80309, USA}
\newcommand{\CUBoulderAddress}{Center for Theory of Quantum Matter, University of Colorado, Boulder, Colorado 80309, USA}

\definecolor{darkgreen}{RGB}{0, 160, 0}
\definecolor{purple}{RGB}{128, 0, 128}
\definecolor{lavender}{RGB}{60, 180, 50}
\definecolor{annoyingpink}{RGB}{255,0,211}
\definecolor{orange}{RGB}{230,85,13}

\begin{document}

\title{Sub-millisecond Entanglement and iSWAP Gate between Molecular Qubits}


\author{Lewis~R.B.~Picard}
\thanks{A.J.P. and L.R.B.P contributed equally to this work.}
\affiliation{\HarvardPhysicsAddress}
\affiliation{\HarvardChemistryaddress} 
\affiliation{\CUAaddress}

\author{Annie~J.~Park}
\thanks{A.J.P. and L.R.B.P contributed equally to this work.}
\affiliation{\HarvardChemistryaddress} 
\affiliation{\HarvardPhysicsAddress}
\affiliation{\CUAaddress}

\author{Gabriel~E.~Patenotte}
\affiliation{\HarvardPhysicsAddress}
\affiliation{\HarvardChemistryaddress} 
\affiliation{\CUAaddress}

\author{Samuel~Gebretsadkan}
\affiliation{\HarvardPhysicsAddress}
\affiliation{\HarvardChemistryaddress} 
\affiliation{\CUAaddress}

\author{David~Wellnitz}
\affiliation{\JILAAddress}
\affiliation{\CUBoulderAddress}

\author{Ana~Maria~Rey}
\affiliation{\JILAAddress}
\affiliation{\CUBoulderAddress}

\author{Kang-Kuen~Ni}
\email{ni@chemistry.harvard.edu}
\affiliation{\HarvardChemistryaddress} 
\affiliation{\HarvardPhysicsAddress}
\affiliation{\CUAaddress}

\date{\today}

\begin{abstract}
Quantum computation (QC) and simulation rely on long-lived qubits with controllable interactions. Early work in quantum computing made use of molecules because of their readily available intramolecular nuclear spin coupling and chemical shifts, along with mature nuclear magnetic resonance techniques~\cite{lloyd_potentially_1993,gershenfeld_bulk_1997}.
Subsequently, the pursuit of many physical platforms has flourished.
Trapped polar molecules have been proposed as a promising quantum computing platform, offering scalability and single-particle addressability while still leveraging inherent complexity and strong couplings of molecules~\cite{DeMille2002, Yelin2006, Zhu2013,Ni2018, Hudson2018}. 
Recent progress in the single quantum state preparation and coherence of the hyperfine-rotational states of individually trapped molecules allows them to serve as promising qubits~\cite{Park_Zwierlein_2017_NaK1sCoherence,gregory_robust_2021,lin2022_NaRb,burchesky_rotational_2021,christakis_probing_2023,park_extended_2023,gregory_second-scale_2024}, with intermolecular dipolar interactions creating entanglement~\cite{Holland_Cheuk_2023_DDI, bao_dipolar_2023}. However, universal two-qubit gates have not been demonstrated with molecules. Here, we harness intrinsic molecular resources to implement a two-qubit iSWAP gate using individually trapped $X^{1}\Sigma^{+}$ NaCs molecules. We characterize the innate dipolar interaction between rotational states and control its strength by tuning the polarization of the traps. By allowing the molecules to interact for 664 $\mu$s at a distance of 1.9 $\mu$m, we create a maximally entangled Bell state with a fidelity of 94(3)\%, following postselection to remove trials with empty traps. Using motion-rotation coupling, we measure residual excitation of the lowest few motional states along the axial trapping direction and find them to be the primary source of decoherence.
Finally, we identify two non-interacting hyperfine states within the ground rotational level in which we encode a qubit. The interaction is toggled by transferring between interacting and non-interacting states to realize an iSWAP gate, which, together with single-qubit rotations, forms a universal set of gates. We verify the gate performance by measuring its logical truth table.

\end{abstract}

\maketitle

\section{Introduction}
Foundational work three decades ago used molecular systems with nuclear magnetic resonances for proof-of-concept quantum computation~\cite{lloyd_potentially_1993,gershenfeld_bulk_1997}, transforming theoretical ideas into early experiments~\cite{jones_implementation_1998,vandersypen_experimental_2001}. These platforms enlisted the accessible complexity
of molecules, using intrinsic nuclear spins as qubits and natural spin-spin interactions to implement two-qubit gates. However, the platform was difficult to scale and lacked true entanglement~\cite{Caves02} because the molecules were dissolved in a solvent at room temperature and subjected to ensemble averaging.
Subsequent developments in quantum computing have shifted toward individually controllable systems with simpler structures, such as trapped ions \cite{monroe_scaling_2013}, neutral atoms \cite{bluvstein_logical_2023}, and superconducting circuits \cite{kjaergaard_superconducting_2020}.
Meanwhile, ultracold trapped polar molecules, which have a complex internal structure spanning vibrational, rotational, and hyperfine manifolds, were proposed as a new kind of qubit \cite{DeMille2002}.
The properties of molecules have been leveraged in recent precision measurements to place orders-of-magnitude improved bounds on the electron electric dipole moment for searches of physics beyond the standard model~\cite{Andreev_ACME_2018_EDM,Roussy_Cornell_2023_EDM}. 
These same molecular resources are also expected to prove valuable in quantum science. In particular, the long-range, tunable dipolar interactions between molecules 
provide a way to perform quantum simulation of exotic quantum phases of matter~\cite{Micheli2006,Gorshkov2011a,Yao2018,Schmidt2022}. 
For quantum information, the dipolar interaction can be used to implement entangling gates and the rich structure of the molecules enables dense encoding of information, which together may offer a new approach to scalability and fault-tolerance \cite{Ni2018, Albert_Preskill_2020_QEC,Sawant_Cornish_2020_Qudit}.

Toward these goals, ultracold polar molecules have been prepared and isolated in optical traps \cite{Ni2008, Danzl2008, Lang2008}.In optical tweezer arrays, control and readout of single quantum states of molecules have been achieved~\cite{Anderegg_Doyle_2019_Tweezer,cairncross_rovibgs_2021,rosenberg_observation_2022,Ruttley_Cornish_2023_Mergoassociation}, and both hyperfine and rotational coherence have been extended well beyond the dipolar interaction timescale~\cite{Park_Zwierlein_2017_NaK1sCoherence,gregory_robust_2021,lin2022_NaRb,burchesky_rotational_2021,christakis_probing_2023,park_extended_2023,gregory_second-scale_2024}. Most recently, entanglement between two isolated polar molecules has been realized~\cite{Holland_Cheuk_2023_DDI,bao_dipolar_2023}.
However, a universal two-qubit logic gate, which could apply to any arbitrary initial state, has not yet been demonstrated. 

In this work we demonstrate a sub-millisecond iSWAP gate, in the class of universal logic gates, to entangle hyperfine states of pairs of individually trapped NaCs molecules. The gate relies on the coherent electric dipole-dipole interaction between rotational states of molecules. We characterize this interaction by creating a two-qubit Bell state within 664 $\mu$s with a measured fidelity of 94(3)\%. The fidelity is limited by motion of molecules along the weakly confined axial dimension of the traps, which we probe using motion-rotation coupling to determine an axial ground state fraction of 32(5)\%. We toggle the dipole-dipole interaction on and off by transferring between rotational and nuclear-spin degrees of freedom, allowing us to encode qubits in a non-interacting subspace. We measure the truth table of the iSWAP gate, with entanglement inferred from the Bell state fidelity. We identify hyperfine-changing pulses as the dominant source of gate infidelity, and offer pathways for near-term improvements. Our results establish multi-level molecules as a viable resources for universal quantum computation and advanced quantum simulations.

\section{Molecular structure}
\label{sec: 2}

\begin{figure}[h]
    \centering
    \includegraphics[width=0.5\columnwidth]{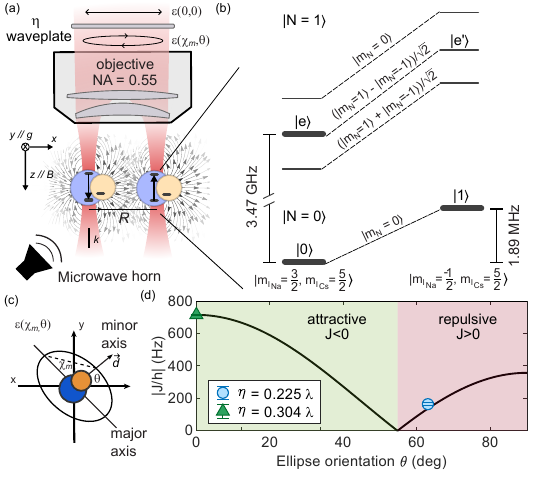}
    \caption{(a) Schematic of the geometry of a pair of molecules trapped in neighbouring 1064 nm optical tweezers. A magnetic field of 864.1 G is applied parallel to the $k$-vector of the tweezer arrays, which is along $z$, and $g$ indicates the direction of gravity. The tweezer polarization $\varepsilon(\chi,\theta)$, defined in the $x-y$ plane by the ellipticity $\chi$ and orientation $\theta$, is tuned to a magic ellipticity condition $\varepsilon(\chi_m,\theta)$ using a waveplate with a retardance of $\eta$ in the path of the initially horizontally polarized light. (b) Relevant internal states are labeled by their dominant hyperfine states in the rotational ground and first rotationally excited manifolds of NaCs. (c) Illustration of the polarization ellipse in the x-y plane, perpendicular to the $k$-vector of the tweezer. (d) Measured interaction strengths $J$ between molecules separated by 1.9 \um~at the magic ellipticity condition for two different choices of waveplate retardance $\eta$. The black curve shows the expected form of the interaction strength scaled to the maximum value of $J/h$ observed at $\eta=0.304~\lambda$.}
    \label{fig:intro}
\end{figure}

Our experiment starts with a pair of individually trapped NaCs molecules
assembled from ground-state-cooled Na and Cs atoms following a procedure detailed previously in refs. \cite{Liu2018,zhang_optical_2022,picard_high_2023}. 
These molecules are held in ``magic'' ellipticity optical tweezers~\cite{Rosenband2018} where the polarization ellipticity is chosen to cancel first-order differential light shifts between the $N=0$  and one sublevel ($\ket{e}$) of the $N=1$ rotational states, yielding long rotational coherence times ~\cite{park_extended_2023}. A schematic of the experimental setup is depicted in Fig. \ref{fig:intro}(a). Three distinct molecular energy levels, labelled $\ket{0}$, $\ket{1}$, and $\ket{e}$, are important for this work and are shown in Fig, \ref{fig:intro}(b). Molecules are initially assembled in $\ket{0}$ within the ground rotational manifold and can be transferred using resonant microwaves to $\ket{e}$ in the first rotationally excited manifold, which couples to $\ket{0}$ via the dipolar interaction. Nuclear spin state $\ket{1}$ in the ground rotational manifold is selected to encode a hyperfine qubit in the molecule (see Appendix for full state labeling). For molecular detection, we sequentially detect $\ket{0}$ and $\ket{e}$ or $\ket{0}$ and $\ket{1}$ within a single experimental sequence by converting molecules back to atoms for fluorescence imaging as described in Ref.~\cite{picard_site-selective_2024}. This multi-state readout is essential as it verifies the presence of a molecule in each tweezer independent of its internal state, allowing us to identify cases where exactly two molecules are detected in the system. The molecule pair production rate is $\sim$3\% of all experimental runs.

A key reason for the choice of NaCs molecules here is their large molecule-frame dipole moment of $d = 4.6$ Debye \cite{Aymar2005}. 
In the absence of an external electric field, the lab-frame permanent dipole moment vanishes. Dipolar interactions then lead to coherent exchange of rotational excitations between pairs of molecules~\cite{Yan2013,christakis_probing_2023,Holland_Cheuk_2023_DDI,bao_dipolar_2023} in the same hyperfine manifold at rate $J$ described by the Hamiltonian \cite{Gorshkov2011a, deLeseleucPRL2017,Ni2018}
\begin{align}
    \hat H_{DD}
    =
    \frac{J}{2}
    \qty(
    \hat s_1^+ \hat s_2^- + \hat s_1^- \hat s_2^+
    )
    =
    \left(\frac{d}{\sqrt{3}}\right)^2\frac{1-3\cos^2{\theta}}{4\pi\epsilon_0R^3}
    \qty(
    \ket{0e} \bra{e0}
    +
    \ket{e0} \bra{0e}
    )\label{eq:Hdd}   
\end{align}
where $R$ is the molecular separation and $\theta$ is the angle between the array axis and the transition dipole moment $d/\sqrt{3}$ of the $\ket{0}\rightarrow\ket{e}$ transition associated with the raising, $\hat s^+_i = \ket e \bra 0_i$, and lowering, $\hat s^-_i = (\hat s^+_i)^\dagger$, operators \cite{Wall2015}.

The dipolar interaction can be tuned by changing $R$ and $\theta$. Uniquely to our system, the rotational eigenstates of molecules here are dominantly quantized by the trapping light rather than static electric or magnetic fields. 
Therefore, the orientation of the tweezer polarization ellipse (Fig.~\ref{fig:intro}(c)) determines the angle $\theta$, providing a way to switch between attractive ($J<0$) and repulsive ($J>0$) interactions. Experimentally, we realize two different $\theta$s by choosing waveplates with a retardance $\eta=0.225 \lambda$ and $0.304 \lambda$ while maintaining the ``magic'' ellipticity polarization ($\chi_m$) condition (Fig. \ref{fig:intro}d). For the remainder of this work, we use the latter waveplate to achieve $\theta=0$, maximizing the magnitude of interaction strength. In the future, the strength could be dynamically tuned using fast polarization control via optical interference of multiple tweezer beams \cite{la_porta_optical_2004,liu_robust_2020}.

\section{Dipole-dipole interaction and entanglement}

\begin{figure*}[ht]
    \centering
    \includegraphics[width=1\columnwidth]{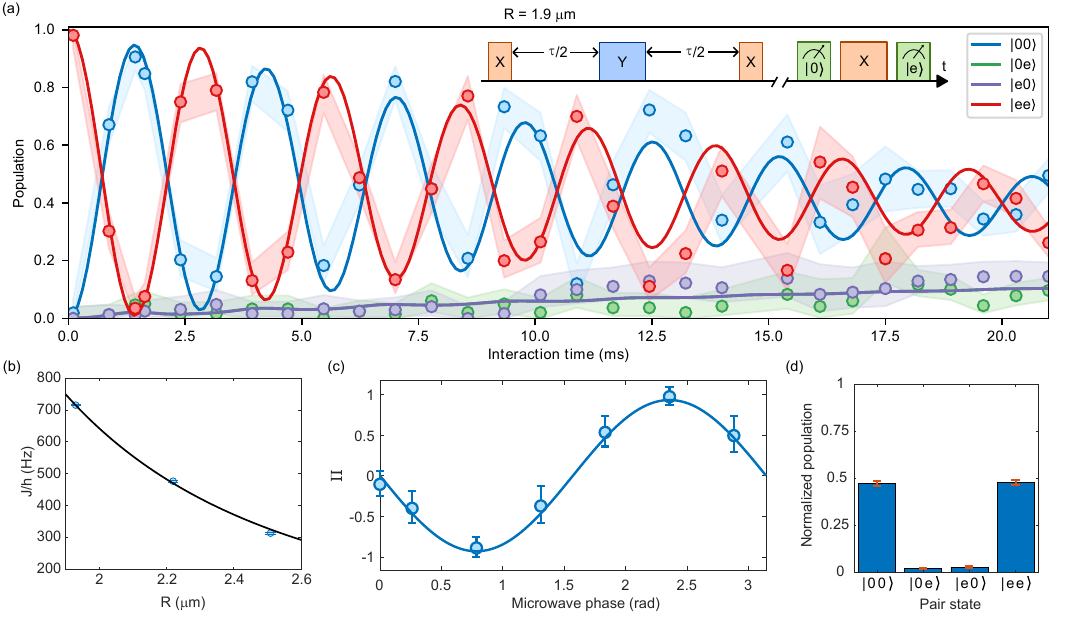}
    \caption{(a) Populations of four possible two-qubit states of a pair of molecules undergoing dipole-dipole mediated rotational exchange at a separation of $R = 1.9~\mu$m, postselected to include only cases where both molecules are detected. The inset shows the spin-echo microwave pulse sequence used to probe the interaction. Solid lines are theory curves from a full quantum-mechanical simulation of the dipolar interaction and molecule motion, with temperature and motion parameters determined from separate single-molecule measurements (see Fig. \ref{fig:motion}), and tweezer separation and single molecule dephasing chosen to match the data. (b) Scaling of the interaction strength $J$ as function of distance, with the solid black line representing a single-parameter fit with the function $a/R^3$ with best fit value $ a = 5.14(4)~\textrm{kHz}~\mu \textrm{m}^{-3}$. (c) Parity oscillation signal $\Pi = P_{00} + P_{ee} - P_{0e} - P_{e0}$ as a function of phase of the global microwave $\pi/2$ pulse applied to the Bell state produced by the dipole-dipole interaction. (d) Corresponding populations of the components of the Bell state prior to applying the global microwave pulse.}
    \label{fig:interactions}
\end{figure*}

Coherent dipolar interaction is the essential ingredient required to entangle molecules with high fidelity. We characterize the coherent exchange of rotational excitation between a pair of molecules 
using a spin echo Ramsey sequence \cite{souza_robust_2012}, illustrated in the inset of Fig. \ref{fig:interactions}(a).
Global microwave pulses are used to address the $\ket{0} \rightarrow \ket{e}$ transition. The dipolar exchange interaction results in an output state with amplitudes determined by the interaction time: $\ket{\psi(\tau)} = \frac{1}{2}(1 - e^{-i\frac{J\tau}{2\hbar}})\ket{00} + \frac{1}{2}(1 + e^{-i\frac{J\tau}{2\hbar}})\ket{ee}$. \\

In Fig. \ref{fig:interactions}(a), we show the evolution of the populations in each of the possible states, $\ket{00}$, $\ket{0e}$, $\ket{e0}$, and $\ket{ee}$, of a pair of molecules separated by 1.9~$\mu$m as a function of the total interaction time~$\tau$. The data are postselected to include only cases in which molecules were detected in both tweezers. 
The populations represent a maximum likelihood estimate of the probability of obtaining each state and the shaded regions are 1$\sigma$ confidence intervals. We analyze the data with the aid of a master equation simulation of the interaction with decoherence from which we extract an interaction rate of $J/h = 715(2)$ Hz (see Appendix). We observe that the decoherence is dominated by noise on the interaction strength, which can be seen from Fig. \ref{fig:interactions}(a) by comparing the symmetric decay of the oscillations between $\ket{00}$ and $\ket{ee}$ with the slower rise of the $\ket{0e}$ and $\ket{e0}$ populations. We model this decoherence using a full quantum mechanical simulation of the motion of the molecules. The solid lines represent the decaying oscillations predicted by this model, with parameters extracted from independent single-molecule measurements shown in Fig. \ref{fig:motion}.
 
We performed the same experiment at separations of 2.2 and 2.5 \um~and extracted $J$ for each distance from the master equation fit (see Extended Data Fig. \ref{fig:distances}). In Fig. \ref{fig:interactions}(b) we plot the interaction rate against the separation and perform a single parameter fit to the function $a/R^3$. We observe good agreement with the expected $1/R^3$ scaling of the dipole-dipole interaction, and extract a distance-independent interaction rate of $\frac{J}{hR^3} = 5.14(4)~\textrm{kHz}~\mu \textrm{m}^{-3}$.
This rate is faster by 17\% than expected from the dipole moment of NaCs~\cite{Aymar2005}, which could be explained by a systematic error of 4\% in the distance calibration. The measured dipolar strength $J$ is the fastest observed in work on ultracold molecules and magnetic atoms, even considering the closer spacings in optical lattice systems~\cite{Yan2013,christakis_probing_2023,chomaz_dipolar_2023}.

The dipole-dipole interaction natively generates entanglement between the rotational states of the molecules \cite{Ni2018}. Applying the spin-echo pulse sequence with an interaction time $h/2J$ maps the input $\ket{00}$ product state to the maximally entangled state $\frac{1}{\sqrt{2}}(\ket{00} - i\ket{ee})$, which is equivalent up to a global rotation to the $\ket{\Phi^\pm}$ Bell states. This was demonstrated experimentally with CaF molecules recently with a fidelity of $<$0.9, benefiting from motional narrowing due the disparate timescales of the dipolar interaction and motion, despite a large motional excited state population~\cite{Holland_Cheuk_2023_DDI,bao_dipolar_2023}. We expect our molecules with a large motional ground-state fraction to do significantly better. We prepare this maximally entangled state at a molecule separation of 1.9 \um, using an interaction time of 664 \us. 
In order to verify the entanglement of the generated state, we measure the off-diagonal coherences of the density matrix by applying an additional global analysis microwave pulse with a variable phase to the pair of molecules. At each phase we measure the parity-signal $\Pi = P_{00} + P_{ee} - P_{0e} - P_{e0}$, which we expect to oscillate with the applied phase $\phi$ as $\Pi = C\sin(2\phi)$, where the amplitude, $C$, of the parity oscillation is the coherence \cite{sackett_experimental_2000}. The total entanglement fidelity is determined by the coherence and the diagonal populations, $F = \frac{1}{2}(C + P_{00} + P_{ee})$. In Fig. \ref{fig:interactions} (c) we show this measured parity oscillation signal, along with the populations of all the two qubit states prior to the analysis pulse in (d). We extract a coherence $C = 0.93(3)$ using a weighted least squares fit, and populations $P_{00} = 0.47(1)$ and $P_{ee} = 0.48(1)$. The total entanglement fidelity, conditioned on the detection of both molecules at the end of the sequence, is thus $F = 0.94(3)$. 
This is consistent to within 1$\sigma$ with a maximum fidelity of 0.97 predicted given the currently observed decoherence of the dipole-dipole interaction.

\section{Motion-rotation coupling}
\label{Section:motion}

\begin{figure}[h]
    \centering
    \includegraphics[width=0.5\columnwidth]{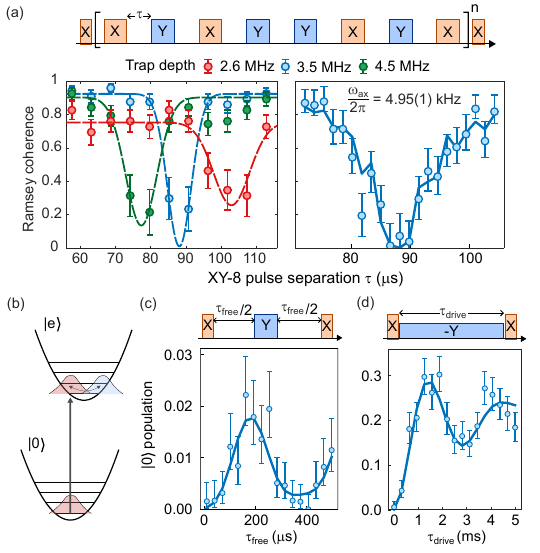}
    \caption{ (a) Measured Ramsey coherence as a function of pulse separation $\tau$ between XY-8 $\pi$-pulses. The left plot shows three different trap depths (inferred from theoretical polarizability of NaCs \cite{Vexiau2017,park_extended_2023}), at which we apply three XY-8 repetitions, totaling 24 $\pi$-pulses over an interrogation time of 24$\tau$. By altering the tweezer depth during the measurement, we can modify the trapping frequency $\omega_{\textrm{ax}}$, which shifts the $\tau$ that produces maximal decoherence. Dashed lines are a Gaussian fit serving as a guide to the eye. The right plot shows a detailed spectrum of the noise peak at the depth of 3.5 MHz at which we perform interactions, probed with six XY-8 repetitions, from which we extract an axial trap frequency of 4.95(1) kHz. (b) Illustration of the motional wavefunction of a single molecule following excitation from $\ket{0}$ to $\ket{e}$ when the $\ket{e}$ trapping potential is displaced. (c) Observed oscillations in $\ket{0}$ at the axial trapping frequency as a function of free evolution time $\tau_{\textrm{free}}$ during spin-echo sequence. (d) Population in $\ket{0}$ under continuous microwave drive along the -Y direction, where the drive Rabi frequency of $\Omega/2\pi = 5.18(3)$ kHz is chosen to be close to the axial trapping frequency. The solid lines in a, c and d are generated by the same theoretical model with temperature, trap displacement, and dephasing parameters chosen to match all of the data simultaneously.}
    \label{fig:motion}
\end{figure}

The dominant dephasing mechanism of the two-body dipolar interaction is noise on the interaction strength. Such noise can be induced by relative motion of interacting molecules, which leads to fluctuation of $R$ and $\theta$ in Equation~\ref{eq:Hdd}. To investigate the effect of motion, we first focus on the rotational coherence between $\ket{0}$ and $\ket{e}$ of a single molecule in a tweezer. While a common way to remove noise and extend single-particle coherence is by applying dynamical decoupling~\cite{christakis_probing_2023,Holland_Cheuk_2023_DDI, bao_dipolar_2023, park_extended_2023}, we found empirically that the coherence is significantly reduced by an XY-8 pulse sequence when the $\pi$-pulse spacing $\tau$ is chosen to decouple noise on the interaction timescale $h/2J$. In particular, we observe a dip in the coherence as a function of $\tau$, the position of which shifts with the trap depth (Fig.~\ref{fig:motion}(a)). We identify this dip as corresponding to the axial trapping frequency of the molecules, originating due to parametric heating of the molecules by pulses spaced by half the trap oscillation period. From a fine scan of this dip, we can extract an axial trap frequency of 4.95(1) kHz at the trap depth at which we perform dipolar interactions.

To explain the parametric heating there has to be a physical coupling between rotation and motional state, implying a deviation from ideal ``magic'' trapping. This could be caused either by a change in the trap depth or a displacement of the trap center between the $\ket{0}$ and $\ket{e}$ states. Because the tweezer polarization is empirically tuned to match the trap depths for the $\ket{0}$ and $\ket{e}$ states \cite{park_extended_2023} we can eliminate the former, and turn towards characterizing the displacement.We construct a minimal model with a state-dependent displacement of the trapping potential as illustrated in Fig.~\ref{fig:motion}(b) described by

\begin{align}
    \hat H_{\rm motion-rotation}
    &=
    \hbar \omega_{\rm ax} \hat a^\dagger \hat a
    -
    \frac{\hbar \zeta \omega_{\rm ax}}{2} 
    \ket e \bra e
    \qty(\hat a + \hat a^\dagger) \, ,
    \label{eq:spin-motion}
\end{align}
and attribute the origin of the coupling to optical aberrations, specifically astigmatism.
Here, $\zeta$ is the displacement in units of harmonic oscillator length $\sqrt{\hbar / (2 m \omega_{\rm ax})} = 80$nm, with $m = 156$a.u.~the mass of NaCs and $\omega_{\rm ax}$ the axial trapping frequency. $\hat a$ is the lowering operator of the axial trap in the harmonic approximation. 

Such motion-rotation coupling can transfer motional excitation to internal rotational states~\cite{gilmore2021quantum}, and vice-versa, allowing us to quantify the various motional effects that contribute to the two-body decoherence seen in Fig.~\ref{fig:interactions}a. In a spin-echo Ramsey experiment (Fig.~\ref{fig:motion}(d) inset), motion will lead to coherent dephasing and re-phasing with half the trap frequency, as the spatial wavepackets of the two internal states $\ket 0$ and $\ket e$ separate and recombine. In harmonic approximation, the functional form is given by
$1 - \exp{\frac{-2\zeta^2 \sin^4(\omega_{\rm ax}t/4)}{\tanh[\hbar \omega_{\rm ax}/(2k_B T)]} }$~\cite{koller2015demagnetization}.
As a consequence, this measurement is  sensitive to the joint parameter $\zeta^2/\tanh[\hbar \omega_{\rm ax}/(k_B T)]$, where $T$ is the molecule temperature and $k_B$ is the Boltzmann constant.

We can couple motion and rotation more directly by applying a continuous drive with a Rabi frequency $\Omega = \omega_{\rm ax}$, which generates dressed rotational states $\ket{\pm} = (\ket{0} \pm \ket e)/\sqrt{2}$ with an energy separation $\hbar\Omega$. In the dressed state picture, the rotation-motion coupling in Eq.~\eqref{eq:spin-motion} enables resonant excitation transfer between motion and rotation driven by $\hat H_\mathrm{eff} = \zeta \omega_{\rm ax}/4\, (\ket{+}\bra{-} \hat a + \ket{-}\bra{+} \hat a^\dagger)$, where rotating terms at frequency $\omega_{\rm ax}$ have been neglected. We prepare the initial state as the lower dressed state $\ket -$ by applying a $\pi/2$ pulse with a phase +90\degree~with respect to the drive such that excitation can initially only be transferred from motion to rotation. A final $\pi/2$ pulse with the same phase maps the dressed states back to the $\ket 0$, $\ket e$ basis for readout. Since this transfer only occurs if initial motional excitation is present, the oscillation amplitude is directly related to the motional ground state fraction. We perform this experiment using a Rabi frequency of $\Omega/2\pi = 5.18(3)$ kHz, slightly detuned from $\omega_{ax}$, 
resulting in oscillation at the generalized motion-dependent Rabi frequency $\sqrt{\hat a^\dagger \hat a\, \zeta^2\omega_{\rm ax}^2/4 + (\Omega - \omega_{\rm ax})^2}$ with decreased amplitude.

To fit the data, we use a discrete variable representation to model the trap generated by a tweezer with astigmatism from first principles (see Appendix). The astigmatism is directly related to the displacement $\zeta$. We compute theory curves for a grid with varying astigmatism, temperature, and motional dephasing, and extract expectation values and error bars from parametric bootstrapping \cite{efron_introduction_1994}. We find $\hbar\omega_{\rm ax} / (k_B T) = 0.39(8)$ corresponding to an axial ground state fraction of 32(5)\%, and $\zeta = 0.062(5)$ from an astigmatism of 0.13(1)$\lambda$.
This ground state fraction is consistent with a previously measured lower bound on the molecule temperature using thermometry of atoms after dissociation \cite{picard_site-selective_2024}, but is lower than expected given the measured Feshbach molecule production efficiency under the assumption of identical trapping frequencies for both atoms ~\cite{Zhang_Ni_2020_MagnetoassociationTweezer}. 

With these parameters, we can model the two-molecule contrast oscillations in Fig.~\ref{fig:interactions}(a). We use the discrete variable representation for single-molecule computations and project dipole interactions into the motional subspace (see Appendix for details). We include single-molecule decoherence with a decoherence time of $\tau = 80$ ms to match the rise of $\ket{e0}$ and $\ket{0e}$ population and set the distance to $1.79$~\um~to match the observed oscillation frequency. We find good quantitative agreement between theory and data. 

For a larger astigmatism of 0.16$\lambda$ at a temperature of $\hbar \omega_{\rm ax} / (k_B T) = 0.4$, our model predicts a Bell state infidelity of 3\%, with a contribution of around 1\% due to single-molecule dephasing (see Appendix). Looking ahead, without single-molecule dephasing and assuming a ground state fraction of 80\%, our model predicts a Bell state infidelity of 0.5\% limited by aberrations. Finally, by eliminating astigmatism, the infidelity could be decreased to $2 \times 10^{-4}$, with approximately equal contributions from residual thermal motion and interaction back-action onto motion~\cite{Chew2022}. This back-action comes from the mechanical force that the dipoles exert on each other, which even at zero temperature can lead to off-resonant population transfer to motionally excited states. This effect can ultimately be reduced by increasing tweezer separation or confinement, to either reduce the force $\sim 1/R^4$ or make population transfer more off-resonant.

\section{Quantum logic gates between hyperfine qubits}
\label{Section:gate}

\begin{figure}
    \centering
    \includegraphics[width=0.5\columnwidth]{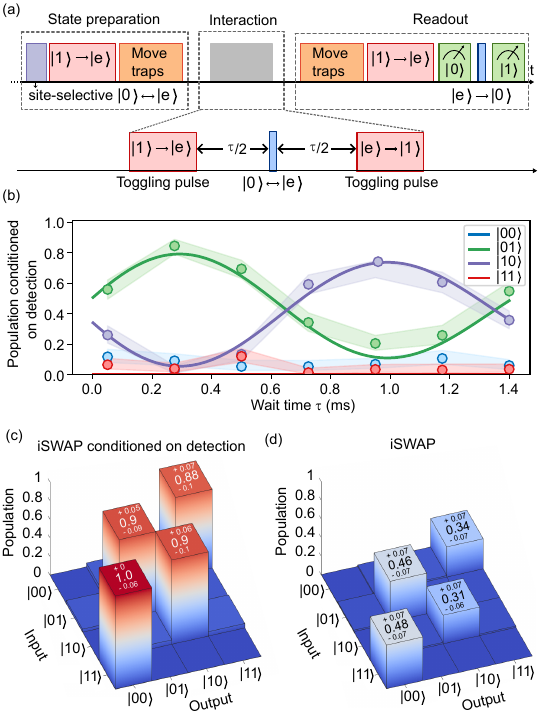}
    \caption{(a) Microwave pulse sequence for implementing an iSWAP gate between two hyperfine qubits. (b) Time-evolution of resonant dipolar exchange for a pair of molecules initially prepared in the two-qubit state $\ket{01}$. Solid lines are theory curves from a master equation model, using parameters extracted from the data in \ref{fig:interactions}. (c) Measured truth tables for the full iSWAP gate sequence. Left shows the probabilities of measuring each two-qubit output state given each input state, relative to the probability of detecting the $\ket{00}$ state with no pulses applied. Losses here are dominated by leakage to the $\ket{e}$ state due to imperfect hyperfine-state changing pulses both during state-preparation and measurement and in the pulses which start and end the gate. Right shows the probabilities postselected to cases where two molecules were detected in the $(\ket{0},\ket{1})$ manifold, which represents the truth table for the entangling interaction with leakage errors caused by the single-qubit gate removed.}
    \label{fig:gate}
\end{figure}

The dipolar exchange interaction can naturally realize an iSWAP gate, which is a universal two-qubit gate, when the duration of the interaction is precisely controlled ~\cite{Ni2018}. 
The Bell state creation protocol demonstrated thus far limited the duration of the interaction using global pulses applied to a particular initial state, but it cannot be generalized to arbitrary initial states and multiple gates. To realize an iSWAP gate for any arbitrary input within the computational basis, we need to switch the dipolar interactions on and off and show the desired logic outcomes. 

Here, we introduce a third state, $\ket{1}$ (Fig.~\ref{fig:intro}b), within the ground rotational manifold, which we use to implement the iSWAP gate with a qubit that does not natively interact~\cite{Ni2018}. The qubit is encoded in a non-interacting hyperfine degree of freedom, which has been shown to be insensitive to external fields and capable of maintaining long coherence times ~\cite{Park_Zwierlein_2017_NaK1sCoherence, gregory_robust_2021, lin2022_NaRb}.  
The state $\ket{1}$ is weakly coupled to $\ket{e}$ by the nuclear quadrupole moment~\cite{Aldegunde2017}, and is chosen among all the accessible hyperfine states because it is the most detuned from $\ket{0}$, allowing for easier individual manipulation. The interaction is toggled by transferring between $\ket{1}$ and the rotationally excited state $\ket{e}$. We note that $\ket{1}$ is strongly coupled to the non-hyperfine changing excited state $\ket{e'}$, which would be an important consideration for the gate implementation, discussed below.

We engineer the interaction between qubits using the pulse sequence shown in Fig.~\ref{fig:gate}(a). We initialize a pair of molecules in the computational basis at a 5~$\mu$m trap separation with site-selective $\ket{0}$ to $\ket{e}$ transfer~\cite{picard_site-selective_2024} and a 310 $\mu$s $\pi$-pulse from $\ket{e}$ to $\ket{1}$ (Extended Data Fig. \ref{fig:hfpulse}). At 1.9 $\mu$m trap separation, we toggle the interaction with the same hyperfine state changing pulse ($\ket{1}\leftrightarrow \ket{e}$), and apply a spin echo pulse on the $\ket{0}\leftrightarrow \ket{e}$ transition in the middle of the interaction period.

For detection, we transfer $\ket{1}$ to $\ket{e}$ at a 5~$\mu$m trap separation and read out the $\ket{0}$ and $\ket{e}$ states. In Fig.~\ref{fig:gate}(b), we show the dipolar interaction between two qubits initialized in the $\ket{01}$ state. Here, the data are postselected for trials where a pair of molecules are both detected within the qubit subspace. The exchange interaction leads to oscillation between the $\ket{01}$ and $\ket{10}$ states. 
The full exchange occurs 330 $\mu$s after the end of the toggling pulse, shorter than if the pulses were infinitely fast, because the interaction occurs already during the duration of the toggling pulse.

At the full exchange time, the sequence realizes the iSWAP gate up to a global rotation, originating from the spin echo pulse. The total gate time, composed of the full exchange time and the duration of both toggling pulses, is 960 $\mu$s. The unitary corresponding to the ideal implementation in the $\left(\ket{0},\ket{1}\right)$ basis is given by 
\begin{equation}
    U = \begin{bmatrix}
        0 & 0 & 0 & -1\\
        0 & -i & 0 & 0\\
        0 & 0 & -i & 0\\
        -1 & 0 & 0 & 0
        \end{bmatrix}
\end{equation}
Under the application of this gate, $\ket{01}$ and $\ket{10}$ obtain a non-trivial phase of $-i$ from the dipolar exchange, which is the same phase we verified via the parity oscillation of the Bell state in Fig.~\ref{fig:interactions}(c). We characterize the logic outcomes of the iSWAP gate for each computational basis state and show the resulting population truth table in Fig.~\ref{fig:gate}(c). For trials where two molecules are detected in the qubit subspace, an average of the probabilities for each input state to transform into the intended output state yields a truth table fidelity~\cite{Hofmann2005_fidelity} of $0.92^{+0.05}_{-0.09}$.

The ease of choosing three states $\ket{0}$, $\ket{1}$, $\ket{e}$ showcases the versatile structure of molecules, but the dense structure also risks causing leakage outside the computational basis~\cite{Ni2018}. 
To take leakage into account, we instead normalize the detected outcomes to the population measured in the initial state $\ket{00}$ before performing site-selective state preparation and the iSWAP gate. The corresponding truth table is shown in Fig.~\ref{fig:gate}(d) and the resulting truth-table fidelity is $0.39^{+0.1}_{-0.09}$.

The current gate fidelity is primarily limited by leakage during four hyperfine changing pulses, which are needed for state preparation, the iSWAP gate, and readout. Rotational transitions that change nuclear spin are two order of magnitude weaker than those that conserve nuclear spin at a magnetic field of 864 G. The strong microwave pulse required to drive the hyperfine-changing transition introduces off-resonant Stark shifts from nearby transitions making the transfer pulse sensitive to microwave amplitude noise. The Stark shifts specifically from non-``magic'' rotational states fluctuate with trap intensity noise and further decohere the transition. Due to these decoherence sources, the current one-way transfer efficiency is limited to 89\%.
In the future, the hyperfine-changing transfer can be made more coherent with active microwave amplitude stabilization and polarization control. The latter also mitigates depopulation of the $\ket{0}$ state, allowing faster driving of the transition. Another source of leakage is due to interactions during the toggling pulses. We anticipate that 9\% of population leaves the computational basis with perfect hyperfine-changing transfer at the current toggling duration. Nevertheless, this leakage may be reduced given the approximately quadratic decrease with pulse duration.
Other methods could reduce such leakage errors, including faster optical or microwave Raman transfers, a lower magnetic field to enhance hyperfine changing coupling for a faster transfer, and moving traps to toggle the interaction. Both leakage sources described here lead to populations remaining $\ket{e}$ and $\ket{e'}$, the latter occurring due to the spin-echo pulse (see Appendix). In future, these populations could be detected without disturbing the computational basis by state-selective transfer to the $N=2$ rotational states for readout. Such detectable errors may be correctable using erasure conversion schemes. 

\section{Conclusions and outlook}
\label{Section:conclusion}

We demonstrate a molecular iSWAP gate, a key requirement for universal quantum computation with polar molecules. The gate is based on the native coherent interaction between molecules, such that noise on the interaction is due more to imperfect state preparation than to variations in external fields. We characterize the interaction by measuring a two-qubit Bell state entanglement fidelity of 94(3)\% with error primarily due to imperfect motional state preparation (32(5)\% axial ground state fraction). A modest improvement in the axial ground state fraction to 80\% and elimination of tweezer astigmatism and single-particle dephasing would allow achieving a fidelity exceeding $0.999$, at which point more fundamental decoherence mechanisms dominate. The iSWAP gate is mainly limited by imperfect transfer between hyperfine states due to the density of nearby molecular states with strong transitions which can induce AC Stark shifts and population leakage. This could be improved through finer microwave polarization control, opening up the complex molecular structure for multi-level applications. Excitations to multiple rotational levels, paired with coherent dipolar interactions, will enable quantum simulation of complex many-body Hamiltonians \cite{Sundar2018,Homeier2024}, and encoding of robust logical qubits \cite{Albert_Preskill_2020_QEC} and qudits \cite{Sawant_Cornish_2020_Qudit}. In parallel, efforts to  pursue efficient molecule production and detection remain important for the utility of the molecular platform, for example, in a hybrid molecule-Rydberg system with orders of magnitude faster gates and mid-circuit readout of molecular states \cite{kuznetsova_rydberg-atom-mediated_2016,wang_enriching_2022,Zhang_Tarbutt_2022_HybridMolRydb,Guttridge2023}. The coherent control of motion-rotation coupling offers a new sensitive probe of molecular temperature and confinement, adding an addional degree of freedom to the molecule toolbox. This promises new opportunities for quantum simulation and metrology based on spin-motion entanglement~\cite{gilmore2021quantum,guardado2021quench,carroll2024observation,scholl2023erasure_motion}. 

\section{Appendix}
\label{sec:methods}

\textbf{External fields and quantization}

We define the quantization axis to be along the magnetic field of 863.9 G, which is kept constant during the experiment and is parallel to the $k$-vector of the optical tweezer. The large magnetic field independently aligns the nuclear spins of the two atoms and decouples them from molecular rotation. The rotational sublevels are split by the AC electric field of the optical tweezer, which is elliptically polarized in the plane perpendicular to the quantization axis. We label three states relevant to this work: the state in which assembled molecules are initially prepared, $\ket{0} \equiv \ket{N=0,m_N=0,m_{I_{\Na}}=3/2,m_{I_{\Cs}}=5/2}$ \cite{cairncross_rovibgs_2021}; the magic state in the first rotationally excited manifold with the same nuclear hyperfine quantum numbers, $\ket{e} \equiv \ket{N=1,m_{I_{\Na}}=3/2,m_{I_{\Cs}}=5/2}\otimes\frac{1}{\sqrt{2}}\left(\ket{m_N=1}-\ket{m_N=-1}\right)$ \cite{park_extended_2023}; and an additional nuclear hyperfine state in the rotational ground manifold that we use as one of our qubit states, $\ket{1} \equiv \ket{N=0,m_N=0,m_{I_{\Na}}=-1/2,m_{I_{\Cs}}=5/2}$. 

\textbf{Molecule preparation and detection}

  The procedure for preparing single ultracold NaCs molecules in optical tweezers is detailed in our previous work \cite{Zhang_Ni_2020_MagnetoassociationTweezer,cairncross_rovibgs_2021,zhang_optical_2022,picard_high_2023}. Briefly, we begin by Raman sideband cooling Na and Cs atoms in separate optical tweezer arrays. We then merge the arrays, magnetoassociate the atoms into molecules at a Feshbach resonance, and perform a two-photon detuned Raman transfer to the rovibrational ground state. Specifically, we prepare the molecules in the  $\ket{0}$ state, which has the largest coupling to the weakly bound Feshbach molecule given our choice of $\sigma^+$ polarization for both the Pump and Stokes lasers used in two-photon transfer. 

After preparing the molecules in an 8-site array, we detect and remove any remaining unassociated Cs atoms and rearrange the molecules to a higher density region at one end of the array. All dipole-dipole interaction data reported in this work includes only the two sites with the highest molecule density at the end of the array. To detect the molecule population, we transfer from $\ket{0}$ back to the weakly bound Feshbach molecule, magnetodissociate  at the Feshbach resonance, and image the resultant Cs atoms. Since the two-photon transfer is far off-resonant for the $\ket{e}$ state, we can perform this process sequentially: first measuring $\ket{0}$ in one image and then transferring remaining population from $\ket{e}$ to $\ket{0}$ before repeating the process. This allows us to read out both $\ket{0}$ and $\ket{e}$ in each shot of the experiment. To initially prepare a pair of molecules in different internal state combinations starting from the $\ket{00}$ state, we employ site-selective excitation of molecules using dynamic ramping of tweezer depths combined with microwave shelving pulses \cite{picard_site-selective_2024}. 

\textbf{Microwave pulse sequences}

To probe dipole-dipole interactions between $\ket{0}$ and $\ket{e}$, we begin with a pair of molecules in the $\ket{00}$ state and apply a $\pi/2$-pulse to prepare the product state $\frac{1}{\sqrt{2}}(\ket{0} + \ket{e})_i\otimes\frac{1}{\sqrt{2}}(\ket{0} + \ket{e})_j$. The phase of this first pulse defines rotation about the x-axis of the Bloch sphere. We then allow the system to evolve under $ H_{DD}$ for time $\tau$ before applying another $\pi/2$ pulse. To cancel out any phase accumulation due to finite detuning of the microwave pulse from resonance or slow drifts in the resonance frequency, we apply a spin-echo $\pi$-pulse in the middle of the interaction period, with a phase of +90\degree~ relative to the initial pulse, realizing an effective y-rotation. This sequence is illustrated in the inset of Fig. \ref{fig:interactions}. The $\pi$-pulses have a total duration of 12.83 \us, and for all pulses addressing this transition we use a truncated Gaussian pulse shape with a width of 0.3 times the pulse duration to reduce off-resonant coupling to other rotational states \cite{boradjiev_control_2013}. To measure the parity oscillation signal for an entangled Bell pair produced using this sequence, we apply an additional $\pi/2$ pulse with a variable phase to the Bell pair.

To investigate the impact of molecule motion, we perform noise-spectroscopy via dynamical decoupling. At a large molecule separation, where there is no significant dipole-dipole interaction on relevant timescales, we perform a Ramsey sequence composed of a $\pi/2$ pulse on $\ket{0} \leftrightarrow \ket{e}$, a series of dynamical decoupling pulses, and a final $\pi/2$ pulse for readout. We use the XY-8 decoupling pulse sequence, consisting of evenly spaced $\pi$-pulses along the x and y axes, illustrated in Fig. \ref{fig:motion}(a)
, which cancels the effect of any pulse area imperfections \cite{souza_robust_2012}. To observe coherent effects of motional dephasing and rephasing we use the same spin-echo pulse sequence as described above, as well as a continuous drive sequence, described in full in the main text.

To perform an iSWAP gate between hyperfine qubits, we toggle the interaction on and off using 310 \us~square $\pi$-pulses on the $\ket{1} \leftrightarrow \ket{e}$ transition. At the midpoint of the interaction period we apply a single spin-echo $\pi$-pulse on $\ket{0} \leftrightarrow \ket{e}$ with a duration of 12.83 \us~and a truncated Gaussian shape, as before.

\textbf{Hyperfine spectroscopy}

To identify a suitable hyperfine state to act as the $\ket{1}$ state of the hyperfine storage qubit, we spectroscopically investigate multiple nuclear-spin-changing transitions from $\ket{e}$. Transitions which change nuclear spin are mediated by the electric quadrupole interaction and are significantly weaker than those which do not. All the identified nuclear spin states are listed in Extended Data Table \ref{tab:HyperfineTransitions}. The state $\ket{1}$ was chosen because it is the state furthest detuned from $\ket{0}$ which still has significant coupling to $\ket{e}$, such that off-resonant coupling to the excited rotational states of the same hyperfine manifold as $\ket{1}$ is suppressed. The full hyperfine structure of NaCs is shown in Extended Data Fig. \ref{fig:hyperfineLevels}, along with a simulated spectrum of the couplings of the $\ket{1}$ and $\ket{e}$ states under a strong microwave drive. 

\textbf{Postselection and error analysis}

To study the interactions between molecules, we postselect on cases where two molecules are detected at the readout stage, which is enabled by the fact that we measure both $\ket{0}$ and $\ket{e}$ during each experimental run. Postselection allows us to eliminate background in the interaction signal due to shots where one or both traps are empty, at the expense of discarding shots in which both molecules were present but one or both were not detected due to measurement error. The only remaining background on the postselected signal is due to the small 0.0038 false positive rate of the imaging step, giving a negligible contribution of $10^{-5}$ to the two-body signal (assuming uncorrelated false positives between sites) \cite{picard_site-selective_2024}. The outcomes of each trial following post-selection are described by a binomial distribution,
$$
p(S|\theta,T)=\begin{pmatrix}T\\S\end{pmatrix}\theta^{S}(1-\theta)^{T-S},
$$
where the number of binomial trials, $T$, is the total number of molecules detected and the number of binomial successes, $S$, is the number detected in a particular state combination with a corresponding population $\theta$. Given a measurement of $S$, the maximum likelihood estimator of the population, $\hat{\theta}$, and associated 1$\sigma$ confidence interval can be computed by inverting the distribution using the Clopper-Pearson method \cite{clopper_use_1934}. 

\textbf{Master equation model of decoherence}

We perform initial analysis of the strength and decoherence of the dipole-dipole interaction using a Lindblad master equation with three phenomenological decay channels corresponding to dephasing of the dipole-dipole interaction itself, relative detuning noise between the two molecules, and global detuning noise on both molecules. These processes are represented, respectively, by the jump operators:
\begin{align}
    L_J = \sqrt{\gamma_J}\left(\ket{0e}\bra{e0} +\ket{e0}\bra{0e}\right) \\
    L_\delta = \sqrt{\gamma_\delta}\left(\ket{e0}\bra{e0} - \ket{0e}\bra{0e}\right) \\
    L_\Delta = \sqrt{\gamma_\Delta}\left(\ket{ee}\bra{ee} - \ket{00}\bra{00}\right)
\end{align}
We fit the master equation model to the data by minimizing the weighted least squares residual of the model and experimental data, using the Nelder-Mead algorithm to optimize the rates of the decoherence processes along with the interaction strength, $J$, and a static relative detuning, $\delta$, between the molecules. We implement the master equation model using the Julia Quantum Optics toolbox \cite{kramer_quantumopticsjl_2018}. 

\textbf{Numerical Simulations of motion} 

To simulate the dynamics with rotation-motion coupling, we use a combination of discrete variable representation (DVR) and exact diagonalization. We compute the local electromagnetic field of a single tweezer in the presence of astigmatism following Ref.~\cite{singh2009tight}. Of the primary aberrations, astigmatism and coma produce displacement of the $\ket{0}$ and $\ket{e}$ states in the axial and radial directions, respectively, by altering the ellipticity of the tweezer polarization at the focus. The local energy of the two states was determined from their polarizabilities $\alpha_0$ and $\alpha_0+\alpha_2\frac{1-3\cos(2\chi)}{10}$, where $\alpha_0$ and $\alpha_2$ are the scalar and tensor polarizabilities of NaCs in a 1064 nm tweezer. We identify the position of minimal energy by gradient descent, which corresponds to the point of maximum intensity for $\ket{0}$.

This gives us a DVR of the potential energy, from which we get the DVR Hamiltonian $\hat H_{1m,0/e}$ for each rotational state following Ref.~\cite{colbert1992novel}.
We numerically compute the $n_{\rm lvl}$ lowest motional eigenenergies and corresponding eigenstates of each rotational state.
We choose a tweezer depth of $h \times 2.45$ MHz to match the numerically extracted level spacing to the experimentally measured value of $h \times 4.95$ kHz.

To compute the two-body Hamiltonian of dimension $4n_{\rm lvl}^2$, we project the dipolar interactions described by the Hamiltonian
\begin{align}
    \hat H_\mathrm{dip}
    &=
    \frac{1}{2}
    \int d^3 \vec r d^3 \vec r'
    \frac{1 - 3\cos^2(\theta)}{4\pi\epsilon_0 \abs{\vec r - \vec r'}^3}
    \qty[
    \hat \psi_e^\dagger(\vec r) \hat \psi_0^\dagger(\vec r')
    \hat \psi_e(\vec r') \hat \psi_0(\vec r)
    + h.c.
    ]
\end{align}
into the low-energy eigenstates, where we replace integrals with sums over discrete grid values.
Analogously, we also project the pulse/drive Hamiltonian
\begin{align}
    \hat H_\mathrm{pulse}
    &=
    \frac{\Omega}{2} \ket{0}\bra{e}
    +
    \frac{\Omega^*}{2} \ket{e}\bra{0}
    +
    \Delta \ket e \bra e
\end{align}
into the DVR basis, which couples different motional levels according to their state overlap.
We choose $\Delta$ such that the microwave pulse is exactly in resonance $\Delta = - \bra{\psi_0} \hat H_{1m,e} \ket{\psi_0} + \bra{\psi_0} \hat H_{1m,0} \ket{\psi_0}$, where $\ket{\psi_0}$ is the ground state of $\hat H_{1m,0}$. We numerically verify that residual detuning $\lesssim h \times 1$ kHz is removed by spin-echo.

Finally, we include single-molecule dephasing at rate $\gamma_\mathrm{deph}$ described by the Lindblad operators
\begin{align}
    \hat L_i
    &=
    \sqrt{\frac{\gamma_\mathrm{deph}}{2}}
    \qty(
    \ket e \bra e_i -  \ket 0 \bra 0_i
    )
    \, .
\end{align}
For the single-molecule simulations, we further include motional dephasing modeled by the diagonal matrix
\begin{align}
    \hat L_{\mathrm{m}, i} &= \sqrt{2\gamma_{\rm m}}
    \begin{pmatrix}
        1 & 0 & \quad \cdots & 0 \\
        0 & 2 & \quad \cdots & 0 \\
        \vdots &\vdots & \quad \ddots & \vdots \\
        0 & 0 & \quad \cdots & n_{\rm lvl}
    \end{pmatrix} \, .
\end{align}

We then propagate the resulting master equation in real time to reproduce the experimental data.
We ignore any dephasing that occurs during the pulses for simplicity, which is expected to be insignificant for pulse durations $\sim 10\mu$s, but keep dipolar interactions on during the pulses.

To make the two-particle simulations more efficient, we ignore off-diagonal elements between motional states whose energies differ by more than $h \times 200$ Hz when computing the lines for Fig.~\ref{fig:interactions}. Since the coupling between these manifolds due to dipolar interactions is $< h \times 10$ Hz, this only affects dynamics during pulses. It removes small oscillations with a few percent amplitude smaller than experimental error bars from the theory curves.

In all simulations, we set $\Omega = h \times 38$ kHz, $\gamma_\mathrm{deph} = 1/$(80 ms), and $R = $ 1.79 $\mu$m, slightly smaller than the experimentally measured value, to match the observed interaction rate for a molecule frame dipole moment $d = 4.6$ D. For the single-molecule simulations we set $n_{\rm lvl} = 20$, for the two-molecule simulations we set $n_{\rm lvl} = 25$.

\textbf{Parametric bootstrapping for fit parameter uncertainty}

To determine the best fit parameters to match the data, we first fix the trap frequency using the XY-8 scan in Fig. \ref{fig:motion}(a) by generating simulated curves for a linear scan of trap frequency values, computing the weighted squared residuals between the curve and the data for each point, and fitting a parabola to determine the best fit trap frequency. To fit the remaining parameters to the XY-8, spin-echo and continuous drive data simultaneously, we generate curves for a 3D grid of the temperature, astigmatism and motional dephasing parameters. We then approximate the best fit value as the grid point which gives the lowest weighted squared residual for the combined data.

To estimate the uncertainties in the optimized parameters of the master equation and motion-rotation models, we apply the parametric bootstrap method \cite{efron_introduction_1994}. For every dataset which we fit, we generate new bootstrap samples using the estimator $\hat{\theta}$ for each two-qubit state at each point, drawing from the distribution $p(S|\hat{\theta},T)$. We then repeat the fitting procedure for each bootstrapped sample, generating a distribution of optimal fit parameters. We generate 300 bootstrap samples for each experimental dataset for the master equation fit, and 500 for the motion-rotation model. The uncertainties in the fit parameters represent the standard deviation of their distribution.

\textbf{Tweezer separation}

We estimate the tweezer separation from the frequencies of the RF tones applied to the acousto-optic deflector used to generate the tweezers and the expected propagation of beams through the 4f optical system \cite{zhang_optical_2022}. This does not account for any systematic deviations of the tweezers from ideal Gaussian beams, such as could be caused by aberrations or by the overlap of the tweezers at small separations. 

\textbf{Data Availability}

The datasets generated during and/or analysed during the current study are available from the corresponding author on reasonable request.

\section{Acknowledgement}
We thank Till Rosenband, Guido Pupillo, Sven Jandura, Matteo Bergonzoni, and Bihui Zhu for discussion. 
This work is supported by AFOSR (FA9550-23-1-0538), NSF (PHY-2110225 \& PFC-PHY-2317149), and AFOSR-MURI (FA9550-20-1-0323 \&  FA9550-21-1-0069).

Note - Ref.~\cite{holland2024demonstration} is a related molecule work  leveraging hyperfine states as qubits. 
\bibliography{master_ref}

\begin{thebibliography}{74}%
\makeatletter
\providecommand \@ifxundefined [1]{%
 \@ifx{#1\undefined}
}%
\providecommand \@ifnum [1]{%
 \ifnum #1\expandafter \@firstoftwo
 \else \expandafter \@secondoftwo
 \fi
}%
\providecommand \@ifx [1]{%
 \ifx #1\expandafter \@firstoftwo
 \else \expandafter \@secondoftwo
 \fi
}%
\providecommand \natexlab [1]{#1}%
\providecommand \enquote  [1]{``#1''}%
\providecommand \bibnamefont  [1]{#1}%
\providecommand \bibfnamefont [1]{#1}%
\providecommand \citenamefont [1]{#1}%
\providecommand \href@noop [0]{\@secondoftwo}%
\providecommand \href [0]{\begingroup \@sanitize@url \@href}%
\providecommand \@href[1]{\@@startlink{#1}\@@href}%
\providecommand \@@href[1]{\endgroup#1\@@endlink}%
\providecommand \@sanitize@url [0]{\catcode `\\12\catcode `\$12\catcode `\&12\catcode `\#12\catcode `\^12\catcode `\_12\catcode `\%12\relax}%
\providecommand \@@startlink[1]{}%
\providecommand \@@endlink[0]{}%
\providecommand \url  [0]{\begingroup\@sanitize@url \@url }%
\providecommand \@url [1]{\endgroup\@href {#1}{\urlprefix }}%
\providecommand \urlprefix  [0]{URL }%
\providecommand \Eprint [0]{\href }%
\providecommand \doibase [0]{https://doi.org/}%
\providecommand \selectlanguage [0]{\@gobble}%
\providecommand \bibinfo  [0]{\@secondoftwo}%
\providecommand \bibfield  [0]{\@secondoftwo}%
\providecommand \translation [1]{[#1]}%
\providecommand \BibitemOpen [0]{}%
\providecommand \bibitemStop [0]{}%
\providecommand \bibitemNoStop [0]{.\EOS\space}%
\providecommand \EOS [0]{\spacefactor3000\relax}%
\providecommand \BibitemShut  [1]{\csname bibitem#1\endcsname}%
\let\auto@bib@innerbib\@empty
\bibitem [{\citenamefont {Lloyd}(1993)}]{lloyd_potentially_1993}%
  \BibitemOpen
  \bibfield  {author} {\bibinfo {author} {\bibfnamefont {S.}~\bibnamefont {Lloyd}},\ }\bibfield  {title} {\bibinfo {title} {A {Potentially} {Realizable} {Quantum} {Computer}},\ }\href {https://doi.org/10.1126/science.261.5128.1569} {\bibfield  {journal} {\bibinfo  {journal} {Science}\ }\textbf {\bibinfo {volume} {261}},\ \bibinfo {pages} {1569} (\bibinfo {year} {1993})}\BibitemShut {NoStop}%
\bibitem [{\citenamefont {Gershenfeld}\ and\ \citenamefont {Chuang}(1997)}]{gershenfeld_bulk_1997}%
  \BibitemOpen
  \bibfield  {author} {\bibinfo {author} {\bibfnamefont {N.~A.}\ \bibnamefont {Gershenfeld}}\ and\ \bibinfo {author} {\bibfnamefont {I.~L.}\ \bibnamefont {Chuang}},\ }\bibfield  {title} {\bibinfo {title} {Bulk {Spin}-{Resonance} {Quantum} {Computation}},\ }\href {https://doi.org/10.1126/science.275.5298.350} {\bibfield  {journal} {\bibinfo  {journal} {Science}\ }\textbf {\bibinfo {volume} {275}},\ \bibinfo {pages} {350} (\bibinfo {year} {1997})}\BibitemShut {NoStop}%
\bibitem [{\citenamefont {DeMille}(2002)}]{DeMille2002}%
  \BibitemOpen
  \bibfield  {author} {\bibinfo {author} {\bibfnamefont {D.}~\bibnamefont {DeMille}},\ }\bibfield  {title} {\bibinfo {title} {Quantum computation with trapped polar molecules},\ }\href {https://doi.org/10.1103/PhysRevLett.88.067901} {\bibfield  {journal} {\bibinfo  {journal} {Phys. Rev. Lett.}\ }\textbf {\bibinfo {volume} {88}},\ \bibinfo {eid} {067901} (\bibinfo {year} {2002})}\BibitemShut {NoStop}%
\bibitem [{\citenamefont {Yelin}\ \emph {et~al.}(2006)\citenamefont {Yelin}, \citenamefont {Kirby},\ and\ \citenamefont {C\^{o}t\'{e}}}]{Yelin2006}%
  \BibitemOpen
  \bibfield  {author} {\bibinfo {author} {\bibfnamefont {S.~F.}\ \bibnamefont {Yelin}}, \bibinfo {author} {\bibfnamefont {K.}~\bibnamefont {Kirby}},\ and\ \bibinfo {author} {\bibfnamefont {R.}~\bibnamefont {C\^{o}t\'{e}}},\ }\bibfield  {title} {\bibinfo {title} {Schemes for robust quantum computation with polar molecules},\ }\href {https://doi.org/10.1103/PhysRevA.74.050301} {\bibfield  {journal} {\bibinfo  {journal} {Phys. Rev. A}\ }\textbf {\bibinfo {volume} {74}},\ \bibinfo {eid} {050301} (\bibinfo {year} {2006})}\BibitemShut {NoStop}%
\bibitem [{\citenamefont {Zhu}\ \emph {et~al.}(2013)\citenamefont {Zhu}, \citenamefont {Kais}, \citenamefont {Wei}, \citenamefont {Herschbach},\ and\ \citenamefont {Friedrich}}]{Zhu2013}%
  \BibitemOpen
  \bibfield  {author} {\bibinfo {author} {\bibfnamefont {J.}~\bibnamefont {Zhu}}, \bibinfo {author} {\bibfnamefont {S.}~\bibnamefont {Kais}}, \bibinfo {author} {\bibfnamefont {Q.}~\bibnamefont {Wei}}, \bibinfo {author} {\bibfnamefont {D.}~\bibnamefont {Herschbach}},\ and\ \bibinfo {author} {\bibfnamefont {B.}~\bibnamefont {Friedrich}},\ }\bibfield  {title} {\bibinfo {title} {{Implementation of quantum logic gates using polar molecules in pendular states}},\ }\href {https://doi.org/10.1063/1.4774058} {\bibfield  {journal} {\bibinfo  {journal} {The Journal of Chemical Physics}\ }\textbf {\bibinfo {volume} {138}},\ \bibinfo {pages} {024104} (\bibinfo {year} {2013})},\ \Eprint {https://arxiv.org/abs/https://pubs.aip.org/aip/jcp/article-pdf/doi/10.1063/1.4774058/14791601/024104\_1\_online.pdf} {https://pubs.aip.org/aip/jcp/article-pdf/doi/10.1063/1.4774058/14791601/024104\_1\_online.pdf} \BibitemShut {NoStop}%
\bibitem [{\citenamefont {Ni}\ \emph {et~al.}(2018)\citenamefont {Ni}, \citenamefont {Rosenband},\ and\ \citenamefont {Grimes}}]{Ni2018}%
  \BibitemOpen
  \bibfield  {author} {\bibinfo {author} {\bibfnamefont {K.-K.}\ \bibnamefont {Ni}}, \bibinfo {author} {\bibfnamefont {T.}~\bibnamefont {Rosenband}},\ and\ \bibinfo {author} {\bibfnamefont {D.~D.}\ \bibnamefont {Grimes}},\ }\bibfield  {title} {\bibinfo {title} {Dipolar exchange quantum logic gate with polar molecules},\ }\href {https://doi.org/10.1039/C8SC02355G} {\bibfield  {journal} {\bibinfo  {journal} {Chem. Sci.}\ }\textbf {\bibinfo {volume} {9}},\ \bibinfo {pages} {6830} (\bibinfo {year} {2018})}\BibitemShut {NoStop}%
\bibitem [{\citenamefont {Hudson}\ and\ \citenamefont {Campbell}(2018)}]{Hudson2018}%
  \BibitemOpen
  \bibfield  {author} {\bibinfo {author} {\bibfnamefont {E.~R.}\ \bibnamefont {Hudson}}\ and\ \bibinfo {author} {\bibfnamefont {W.~C.}\ \bibnamefont {Campbell}},\ }\bibfield  {title} {\bibinfo {title} {Dipolar quantum logic for freely rotating trapped molecular ions},\ }\href {https://doi.org/10.1103/PhysRevA.98.040302} {\bibfield  {journal} {\bibinfo  {journal} {Phys. Rev. A}\ }\textbf {\bibinfo {volume} {98}},\ \bibinfo {pages} {040302(R)} (\bibinfo {year} {2018})}\BibitemShut {NoStop}%
\bibitem [{\citenamefont {Park}\ \emph {et~al.}(2017)\citenamefont {Park}, \citenamefont {Yan}, \citenamefont {Loh}, \citenamefont {Will},\ and\ \citenamefont {Zwierlein}}]{Park_Zwierlein_2017_NaK1sCoherence}%
  \BibitemOpen
  \bibfield  {author} {\bibinfo {author} {\bibfnamefont {J.~W.}\ \bibnamefont {Park}}, \bibinfo {author} {\bibfnamefont {Z.~Z.}\ \bibnamefont {Yan}}, \bibinfo {author} {\bibfnamefont {H.}~\bibnamefont {Loh}}, \bibinfo {author} {\bibfnamefont {S.~A.}\ \bibnamefont {Will}},\ and\ \bibinfo {author} {\bibfnamefont {M.~W.}\ \bibnamefont {Zwierlein}},\ }\bibfield  {title} {\bibinfo {title} {Second-scale nuclear spin coherence time of ultracold {{}$^{23}$Na$^{40}$K} molecules},\ }\href {https://doi.org/10.1126/science.aal5066} {\bibfield  {journal} {\bibinfo  {journal} {Science}\ }\textbf {\bibinfo {volume} {357}},\ \bibinfo {pages} {372} (\bibinfo {year} {2017})}\BibitemShut {NoStop}%
\bibitem [{\citenamefont {Gregory}\ \emph {et~al.}(2021)\citenamefont {Gregory}, \citenamefont {Blackmore}, \citenamefont {Bromley}, \citenamefont {Hutson},\ and\ \citenamefont {Cornish}}]{gregory_robust_2021}%
  \BibitemOpen
  \bibfield  {author} {\bibinfo {author} {\bibfnamefont {P.~D.}\ \bibnamefont {Gregory}}, \bibinfo {author} {\bibfnamefont {J.~A.}\ \bibnamefont {Blackmore}}, \bibinfo {author} {\bibfnamefont {S.~L.}\ \bibnamefont {Bromley}}, \bibinfo {author} {\bibfnamefont {J.~M.}\ \bibnamefont {Hutson}},\ and\ \bibinfo {author} {\bibfnamefont {S.~L.}\ \bibnamefont {Cornish}},\ }\bibfield  {title} {\bibinfo {title} {Robust storage qubits in ultracold polar molecules},\ }\href {https://doi.org/10.1038/s41567-021-01328-7} {\bibfield  {journal} {\bibinfo  {journal} {Nature Physics}\ }\textbf {\bibinfo {volume} {17}},\ \bibinfo {pages} {1149} (\bibinfo {year} {2021})}\BibitemShut {NoStop}%
\bibitem [{\citenamefont {Lin}\ \emph {et~al.}(2022)\citenamefont {Lin}, \citenamefont {He}, \citenamefont {Jin}, \citenamefont {Chen},\ and\ \citenamefont {Wang}}]{lin2022_NaRb}%
  \BibitemOpen
  \bibfield  {author} {\bibinfo {author} {\bibfnamefont {J.}~\bibnamefont {Lin}}, \bibinfo {author} {\bibfnamefont {J.}~\bibnamefont {He}}, \bibinfo {author} {\bibfnamefont {M.}~\bibnamefont {Jin}}, \bibinfo {author} {\bibfnamefont {G.}~\bibnamefont {Chen}},\ and\ \bibinfo {author} {\bibfnamefont {D.}~\bibnamefont {Wang}},\ }\bibfield  {title} {\bibinfo {title} {Seconds-scale coherence on nuclear spin transitions of ultracold polar molecules in 3d optical lattices},\ }\href {https://doi.org/10.1103/PhysRevLett.128.223201} {\bibfield  {journal} {\bibinfo  {journal} {Phys. Rev. Lett.}\ }\textbf {\bibinfo {volume} {128}},\ \bibinfo {pages} {223201} (\bibinfo {year} {2022})}\BibitemShut {NoStop}%
\bibitem [{\citenamefont {Burchesky}\ \emph {et~al.}(2021)\citenamefont {Burchesky}, \citenamefont {Anderegg}, \citenamefont {Bao}, \citenamefont {Yu}, \citenamefont {Chae}, \citenamefont {Ketterle}, \citenamefont {Ni},\ and\ \citenamefont {Doyle}}]{burchesky_rotational_2021}%
  \BibitemOpen
  \bibfield  {author} {\bibinfo {author} {\bibfnamefont {S.}~\bibnamefont {Burchesky}}, \bibinfo {author} {\bibfnamefont {L.}~\bibnamefont {Anderegg}}, \bibinfo {author} {\bibfnamefont {Y.}~\bibnamefont {Bao}}, \bibinfo {author} {\bibfnamefont {S.~S.}\ \bibnamefont {Yu}}, \bibinfo {author} {\bibfnamefont {E.}~\bibnamefont {Chae}}, \bibinfo {author} {\bibfnamefont {W.}~\bibnamefont {Ketterle}}, \bibinfo {author} {\bibfnamefont {K.-K.}\ \bibnamefont {Ni}},\ and\ \bibinfo {author} {\bibfnamefont {J.~M.}\ \bibnamefont {Doyle}},\ }\bibfield  {title} {\bibinfo {title} {Rotational {Coherence} {Times} of {Polar} {Molecules} in {Optical} {Tweezers}},\ }\href {https://doi.org/10.1103/PhysRevLett.127.123202} {\bibfield  {journal} {\bibinfo  {journal} {Physical Review Letters}\ }\textbf {\bibinfo {volume} {127}},\ \bibinfo {pages} {123202} (\bibinfo {year} {2021})}\BibitemShut {NoStop}%
\bibitem [{\citenamefont {Christakis}\ \emph {et~al.}(2023)\citenamefont {Christakis}, \citenamefont {Rosenberg}, \citenamefont {Raj}, \citenamefont {Chi}, \citenamefont {Morningstar}, \citenamefont {Huse}, \citenamefont {Yan},\ and\ \citenamefont {Bakr}}]{christakis_probing_2023}%
  \BibitemOpen
  \bibfield  {author} {\bibinfo {author} {\bibfnamefont {L.}~\bibnamefont {Christakis}}, \bibinfo {author} {\bibfnamefont {J.~S.}\ \bibnamefont {Rosenberg}}, \bibinfo {author} {\bibfnamefont {R.}~\bibnamefont {Raj}}, \bibinfo {author} {\bibfnamefont {S.}~\bibnamefont {Chi}}, \bibinfo {author} {\bibfnamefont {A.}~\bibnamefont {Morningstar}}, \bibinfo {author} {\bibfnamefont {D.~A.}\ \bibnamefont {Huse}}, \bibinfo {author} {\bibfnamefont {Z.~Z.}\ \bibnamefont {Yan}},\ and\ \bibinfo {author} {\bibfnamefont {W.~S.}\ \bibnamefont {Bakr}},\ }\bibfield  {title} {\bibinfo {title} {Probing site-resolved correlations in a spin system of ultracold molecules},\ }\href {https://www.nature.com/articles/s41586-022-05558-4} {\bibfield  {journal} {\bibinfo  {journal} {Nature}\ }\textbf {\bibinfo {volume} {614}} (\bibinfo {year} {2023})}\BibitemShut {NoStop}%
\bibitem [{\citenamefont {Park}\ \emph {et~al.}(2023)\citenamefont {Park}, \citenamefont {Picard}, \citenamefont {Patenotte}, \citenamefont {Zhang}, \citenamefont {Rosenband},\ and\ \citenamefont {Ni}}]{park_extended_2023}%
  \BibitemOpen
  \bibfield  {author} {\bibinfo {author} {\bibfnamefont {A.~J.}\ \bibnamefont {Park}}, \bibinfo {author} {\bibfnamefont {L.~R.}\ \bibnamefont {Picard}}, \bibinfo {author} {\bibfnamefont {G.~E.}\ \bibnamefont {Patenotte}}, \bibinfo {author} {\bibfnamefont {J.~T.}\ \bibnamefont {Zhang}}, \bibinfo {author} {\bibfnamefont {T.}~\bibnamefont {Rosenband}},\ and\ \bibinfo {author} {\bibfnamefont {K.-K.}\ \bibnamefont {Ni}},\ }\bibfield  {title} {\bibinfo {title} {Extended {Rotational} {Coherence} of {Polar} {Molecules} in an {Elliptically} {Polarized} {Trap}},\ }\href {https://doi.org/10.1103/PhysRevLett.131.183401} {\bibfield  {journal} {\bibinfo  {journal} {Phys. Rev. Lett.}\ }\textbf {\bibinfo {volume} {131}},\ \bibinfo {pages} {183401} (\bibinfo {year} {2023})}\BibitemShut {NoStop}%
\bibitem [{\citenamefont {Gregory}\ \emph {et~al.}(2024)\citenamefont {Gregory}, \citenamefont {Fernley}, \citenamefont {Tao}, \citenamefont {Bromley}, \citenamefont {Stepp}, \citenamefont {Zhang}, \citenamefont {Kotochigova}, \citenamefont {Hazzard},\ and\ \citenamefont {Cornish}}]{gregory_second-scale_2024}%
  \BibitemOpen
  \bibfield  {author} {\bibinfo {author} {\bibfnamefont {P.~D.}\ \bibnamefont {Gregory}}, \bibinfo {author} {\bibfnamefont {L.~M.}\ \bibnamefont {Fernley}}, \bibinfo {author} {\bibfnamefont {A.~L.}\ \bibnamefont {Tao}}, \bibinfo {author} {\bibfnamefont {S.~L.}\ \bibnamefont {Bromley}}, \bibinfo {author} {\bibfnamefont {J.}~\bibnamefont {Stepp}}, \bibinfo {author} {\bibfnamefont {Z.}~\bibnamefont {Zhang}}, \bibinfo {author} {\bibfnamefont {S.}~\bibnamefont {Kotochigova}}, \bibinfo {author} {\bibfnamefont {K.~R.~A.}\ \bibnamefont {Hazzard}},\ and\ \bibinfo {author} {\bibfnamefont {S.~L.}\ \bibnamefont {Cornish}},\ }\bibfield  {title} {\bibinfo {title} {Second-scale rotational coherence and dipolar interactions in a gas of ultracold polar molecules},\ }\bibfield  {journal} {\bibinfo  {journal} {Nature Physics}\ }\href {https://doi.org/10.1038/s41567-023-02328-5} {10.1038/s41567-023-02328-5} (\bibinfo {year} {2024})\BibitemShut {NoStop}%
\bibitem [{\citenamefont {Holland}\ \emph {et~al.}(2023)\citenamefont {Holland}, \citenamefont {Lu},\ and\ \citenamefont {Cheuk}}]{Holland_Cheuk_2023_DDI}%
  \BibitemOpen
  \bibfield  {author} {\bibinfo {author} {\bibfnamefont {C.~M.}\ \bibnamefont {Holland}}, \bibinfo {author} {\bibfnamefont {Y.}~\bibnamefont {Lu}},\ and\ \bibinfo {author} {\bibfnamefont {L.~W.}\ \bibnamefont {Cheuk}},\ }\bibfield  {title} {\bibinfo {title} {On-demand entanglement of molecules in a reconfigurable optical tweezer array},\ }\href {https://doi.org/10.1126/science.adf4272} {\bibfield  {journal} {\bibinfo  {journal} {Science}\ }\textbf {\bibinfo {volume} {382}},\ \bibinfo {pages} {1143} (\bibinfo {year} {2023})}\BibitemShut {NoStop}%
\bibitem [{\citenamefont {Bao}\ \emph {et~al.}(2023)\citenamefont {Bao}, \citenamefont {Yu}, \citenamefont {Anderegg}, \citenamefont {Chae}, \citenamefont {Ketterle}, \citenamefont {Ni},\ and\ \citenamefont {Doyle}}]{bao_dipolar_2023}%
  \BibitemOpen
  \bibfield  {author} {\bibinfo {author} {\bibfnamefont {Y.}~\bibnamefont {Bao}}, \bibinfo {author} {\bibfnamefont {S.~S.}\ \bibnamefont {Yu}}, \bibinfo {author} {\bibfnamefont {L.}~\bibnamefont {Anderegg}}, \bibinfo {author} {\bibfnamefont {E.}~\bibnamefont {Chae}}, \bibinfo {author} {\bibfnamefont {W.}~\bibnamefont {Ketterle}}, \bibinfo {author} {\bibfnamefont {K.-K.}\ \bibnamefont {Ni}},\ and\ \bibinfo {author} {\bibfnamefont {J.~M.}\ \bibnamefont {Doyle}},\ }\bibfield  {title} {\bibinfo {title} {Dipolar spin-exchange and entanglement between molecules in an optical tweezer array},\ }\href {https://doi.org/10.1126/science.adf8999} {\bibfield  {journal} {\bibinfo  {journal} {Science}\ }\textbf {\bibinfo {volume} {382}},\ \bibinfo {pages} {1138} (\bibinfo {year} {2023})}\BibitemShut {NoStop}%
\bibitem [{\citenamefont {Jones}\ and\ \citenamefont {Mosca}(1998)}]{jones_implementation_1998}%
  \BibitemOpen
  \bibfield  {author} {\bibinfo {author} {\bibfnamefont {J.~A.}\ \bibnamefont {Jones}}\ and\ \bibinfo {author} {\bibfnamefont {M.}~\bibnamefont {Mosca}},\ }\bibfield  {title} {\bibinfo {title} {Implementation of a quantum algorithm on a nuclear magnetic resonance quantum computer},\ }\href {https://doi.org/10.1063/1.476739} {\bibfield  {journal} {\bibinfo  {journal} {The Journal of Chemical Physics}\ }\textbf {\bibinfo {volume} {109}},\ \bibinfo {pages} {1648} (\bibinfo {year} {1998})}\BibitemShut {NoStop}%
\bibitem [{\citenamefont {Vandersypen}\ \emph {et~al.}(2001)\citenamefont {Vandersypen}, \citenamefont {Steffen}, \citenamefont {Breyta}, \citenamefont {Yannoni}, \citenamefont {Sherwood},\ and\ \citenamefont {Chuang}}]{vandersypen_experimental_2001}%
  \BibitemOpen
  \bibfield  {author} {\bibinfo {author} {\bibfnamefont {L.~M.~K.}\ \bibnamefont {Vandersypen}}, \bibinfo {author} {\bibfnamefont {M.}~\bibnamefont {Steffen}}, \bibinfo {author} {\bibfnamefont {G.}~\bibnamefont {Breyta}}, \bibinfo {author} {\bibfnamefont {C.~S.}\ \bibnamefont {Yannoni}}, \bibinfo {author} {\bibfnamefont {M.~H.}\ \bibnamefont {Sherwood}},\ and\ \bibinfo {author} {\bibfnamefont {I.~L.}\ \bibnamefont {Chuang}},\ }\bibfield  {title} {\bibinfo {title} {Experimental realization of {Shor}'s quantum factoring algorithm using nuclear magnetic resonance},\ }\href {https://doi.org/10.1038/414883a} {\bibfield  {journal} {\bibinfo  {journal} {Nature}\ }\textbf {\bibinfo {volume} {414}},\ \bibinfo {pages} {883} (\bibinfo {year} {2001})}\BibitemShut {NoStop}%
\bibitem [{\citenamefont {Menicucci}\ and\ \citenamefont {Caves}(2002)}]{Caves02}%
  \BibitemOpen
  \bibfield  {author} {\bibinfo {author} {\bibfnamefont {N.~C.}\ \bibnamefont {Menicucci}}\ and\ \bibinfo {author} {\bibfnamefont {C.~M.}\ \bibnamefont {Caves}},\ }\bibfield  {title} {\bibinfo {title} {Local realistic model for the dynamics of bulk-ensemble nmr information processing},\ }\href {https://doi.org/10.1103/PhysRevLett.88.167901} {\bibfield  {journal} {\bibinfo  {journal} {Phys. Rev. Lett.}\ }\textbf {\bibinfo {volume} {88}},\ \bibinfo {pages} {167901} (\bibinfo {year} {2002})}\BibitemShut {NoStop}%
\bibitem [{\citenamefont {Monroe}\ and\ \citenamefont {Kim}(2013)}]{monroe_scaling_2013}%
  \BibitemOpen
  \bibfield  {author} {\bibinfo {author} {\bibfnamefont {C.}~\bibnamefont {Monroe}}\ and\ \bibinfo {author} {\bibfnamefont {J.}~\bibnamefont {Kim}},\ }\bibfield  {title} {\bibinfo {title} {Scaling the {Ion} {Trap} {Quantum} {Processor}},\ }\href {https://doi.org/10.1126/science.1231298} {\bibfield  {journal} {\bibinfo  {journal} {Science}\ }\textbf {\bibinfo {volume} {339}},\ \bibinfo {pages} {1164} (\bibinfo {year} {2013})}\BibitemShut {NoStop}%
\bibitem [{\citenamefont {Bluvstein}\ \emph {et~al.}(2023)\citenamefont {Bluvstein}, \citenamefont {Evered}, \citenamefont {Geim}, \citenamefont {Li}, \citenamefont {Zhou}, \citenamefont {Manovitz}, \citenamefont {Ebadi}, \citenamefont {Cain}, \citenamefont {Kalinowski}, \citenamefont {Hangleiter}, \citenamefont {Ataides}, \citenamefont {Maskara}, \citenamefont {Cong}, \citenamefont {Gao}, \citenamefont {Rodriguez}, \citenamefont {Karolyshyn}, \citenamefont {Semeghini}, \citenamefont {Gullans}, \citenamefont {Greiner}, \citenamefont {Vuletić},\ and\ \citenamefont {Lukin}}]{bluvstein_logical_2023}%
  \BibitemOpen
  \bibfield  {author} {\bibinfo {author} {\bibfnamefont {D.}~\bibnamefont {Bluvstein}}, \bibinfo {author} {\bibfnamefont {S.~J.}\ \bibnamefont {Evered}}, \bibinfo {author} {\bibfnamefont {A.~A.}\ \bibnamefont {Geim}}, \bibinfo {author} {\bibfnamefont {S.~H.}\ \bibnamefont {Li}}, \bibinfo {author} {\bibfnamefont {H.}~\bibnamefont {Zhou}}, \bibinfo {author} {\bibfnamefont {T.}~\bibnamefont {Manovitz}}, \bibinfo {author} {\bibfnamefont {S.}~\bibnamefont {Ebadi}}, \bibinfo {author} {\bibfnamefont {M.}~\bibnamefont {Cain}}, \bibinfo {author} {\bibfnamefont {M.}~\bibnamefont {Kalinowski}}, \bibinfo {author} {\bibfnamefont {D.}~\bibnamefont {Hangleiter}}, \bibinfo {author} {\bibfnamefont {J.~P.~B.}\ \bibnamefont {Ataides}}, \bibinfo {author} {\bibfnamefont {N.}~\bibnamefont {Maskara}}, \bibinfo {author} {\bibfnamefont {I.}~\bibnamefont {Cong}}, \bibinfo {author} {\bibfnamefont {X.}~\bibnamefont {Gao}}, \bibinfo {author} {\bibfnamefont {P.~S.}\ \bibnamefont {Rodriguez}}, \bibinfo {author} {\bibfnamefont
  {T.}~\bibnamefont {Karolyshyn}}, \bibinfo {author} {\bibfnamefont {G.}~\bibnamefont {Semeghini}}, \bibinfo {author} {\bibfnamefont {M.~J.}\ \bibnamefont {Gullans}}, \bibinfo {author} {\bibfnamefont {M.}~\bibnamefont {Greiner}}, \bibinfo {author} {\bibfnamefont {V.}~\bibnamefont {Vuletić}},\ and\ \bibinfo {author} {\bibfnamefont {M.~D.}\ \bibnamefont {Lukin}},\ }\bibfield  {title} {\bibinfo {title} {Logical quantum processor based on reconfigurable atom arrays},\ }\href {https://doi.org/10.1038/s41586-023-06927-3} {\bibfield  {journal} {\bibinfo  {journal} {Nature}\ }\textbf {\bibinfo {volume} {624}},\ \bibinfo {pages} {1} (\bibinfo {year} {2023})}\BibitemShut {NoStop}%
\bibitem [{\citenamefont {Kjaergaard}\ \emph {et~al.}(2020)\citenamefont {Kjaergaard}, \citenamefont {Schwartz}, \citenamefont {Braumüller}, \citenamefont {Krantz}, \citenamefont {Wang}, \citenamefont {Gustavsson},\ and\ \citenamefont {Oliver}}]{kjaergaard_superconducting_2020}%
  \BibitemOpen
  \bibfield  {author} {\bibinfo {author} {\bibfnamefont {M.}~\bibnamefont {Kjaergaard}}, \bibinfo {author} {\bibfnamefont {M.~E.}\ \bibnamefont {Schwartz}}, \bibinfo {author} {\bibfnamefont {J.}~\bibnamefont {Braumüller}}, \bibinfo {author} {\bibfnamefont {P.}~\bibnamefont {Krantz}}, \bibinfo {author} {\bibfnamefont {J.~I.-J.}\ \bibnamefont {Wang}}, \bibinfo {author} {\bibfnamefont {S.}~\bibnamefont {Gustavsson}},\ and\ \bibinfo {author} {\bibfnamefont {W.~D.}\ \bibnamefont {Oliver}},\ }\bibfield  {title} {\bibinfo {title} {Superconducting {Qubits}: {Current} {State} of {Play}},\ }\href {https://doi.org/10.1146/annurev-conmatphys-031119-050605} {\bibfield  {journal} {\bibinfo  {journal} {Annual Review of Condensed Matter Physics}\ }\textbf {\bibinfo {volume} {11}},\ \bibinfo {pages} {369} (\bibinfo {year} {2020})}\BibitemShut {NoStop}%
\bibitem [{\citenamefont {Andreev}\ \emph {et~al.}(2018)\citenamefont {Andreev}, \citenamefont {Ang}, \citenamefont {DeMille}, \citenamefont {Doyle}, \citenamefont {Gabrielse}, \citenamefont {Haefner}, \citenamefont {Hutzler}, \citenamefont {Lasner}, \citenamefont {Meisenhelder}, \citenamefont {O’Leary}, \citenamefont {Panda}, \citenamefont {West}, \citenamefont {West},\ and\ \citenamefont {Wu}}]{Andreev_ACME_2018_EDM}%
  \BibitemOpen
  \bibfield  {author} {\bibinfo {author} {\bibfnamefont {V.}~\bibnamefont {Andreev}}, \bibinfo {author} {\bibfnamefont {D.~G.}\ \bibnamefont {Ang}}, \bibinfo {author} {\bibfnamefont {D.}~\bibnamefont {DeMille}}, \bibinfo {author} {\bibfnamefont {J.~M.}\ \bibnamefont {Doyle}}, \bibinfo {author} {\bibfnamefont {G.}~\bibnamefont {Gabrielse}}, \bibinfo {author} {\bibfnamefont {J.}~\bibnamefont {Haefner}}, \bibinfo {author} {\bibfnamefont {N.~R.}\ \bibnamefont {Hutzler}}, \bibinfo {author} {\bibfnamefont {Z.}~\bibnamefont {Lasner}}, \bibinfo {author} {\bibfnamefont {C.}~\bibnamefont {Meisenhelder}}, \bibinfo {author} {\bibfnamefont {B.~R.}\ \bibnamefont {O’Leary}}, \bibinfo {author} {\bibfnamefont {C.~D.}\ \bibnamefont {Panda}}, \bibinfo {author} {\bibfnamefont {A.~D.}\ \bibnamefont {West}}, \bibinfo {author} {\bibfnamefont {E.~P.}\ \bibnamefont {West}},\ and\ \bibinfo {author} {\bibfnamefont {X.}~\bibnamefont {Wu}},\ }\bibfield  {title} {\bibinfo {title} {{Improved limit on the electric dipole moment of the
  electron}},\ }\href {https://doi.org/10.1038/s41586-018-0599-8} {\bibfield  {journal} {\bibinfo  {journal} {Nature}\ }\textbf {\bibinfo {volume} {562}},\ \bibinfo {pages} {355} (\bibinfo {year} {2018})}\BibitemShut {NoStop}%
\bibitem [{\citenamefont {Roussy}\ \emph {et~al.}(2023)\citenamefont {Roussy}, \citenamefont {Caldwell}, \citenamefont {Wright}, \citenamefont {Cairncross}, \citenamefont {Shagam}, \citenamefont {Ng}, \citenamefont {Schlossberger}, \citenamefont {Park}, \citenamefont {Wang}, \citenamefont {Ye},\ and\ \citenamefont {Cornell}}]{Roussy_Cornell_2023_EDM}%
  \BibitemOpen
  \bibfield  {author} {\bibinfo {author} {\bibfnamefont {T.~S.}\ \bibnamefont {Roussy}}, \bibinfo {author} {\bibfnamefont {L.}~\bibnamefont {Caldwell}}, \bibinfo {author} {\bibfnamefont {T.}~\bibnamefont {Wright}}, \bibinfo {author} {\bibfnamefont {W.~B.}\ \bibnamefont {Cairncross}}, \bibinfo {author} {\bibfnamefont {Y.}~\bibnamefont {Shagam}}, \bibinfo {author} {\bibfnamefont {K.~B.}\ \bibnamefont {Ng}}, \bibinfo {author} {\bibfnamefont {N.}~\bibnamefont {Schlossberger}}, \bibinfo {author} {\bibfnamefont {S.~Y.}\ \bibnamefont {Park}}, \bibinfo {author} {\bibfnamefont {A.}~\bibnamefont {Wang}}, \bibinfo {author} {\bibfnamefont {J.}~\bibnamefont {Ye}},\ and\ \bibinfo {author} {\bibfnamefont {E.~A.}\ \bibnamefont {Cornell}},\ }\bibfield  {title} {\bibinfo {title} {An improved bound on the electron’s electric dipole moment},\ }\href {https://doi.org/10.1126/science.adg4084} {\bibfield  {journal} {\bibinfo  {journal} {Science}\ }\textbf {\bibinfo {volume} {381}},\ \bibinfo {pages} {46} (\bibinfo {year}
  {2023})}\BibitemShut {NoStop}%
\bibitem [{\citenamefont {Micheli}\ \emph {et~al.}(2006)\citenamefont {Micheli}, \citenamefont {Brennen},\ and\ \citenamefont {Zoller}}]{Micheli2006}%
  \BibitemOpen
  \bibfield  {author} {\bibinfo {author} {\bibfnamefont {A.}~\bibnamefont {Micheli}}, \bibinfo {author} {\bibfnamefont {G.}~\bibnamefont {Brennen}},\ and\ \bibinfo {author} {\bibfnamefont {P.}~\bibnamefont {Zoller}},\ }\bibfield  {title} {\bibinfo {title} {A toolbox for lattice-spin models with polar molecules},\ }\href {https://www.nature.com/articles/nphys287} {\bibfield  {journal} {\bibinfo  {journal} {Nat. Phys.}\ }\textbf {\bibinfo {volume} {2}},\ \bibinfo {pages} {341} (\bibinfo {year} {2006})}\BibitemShut {NoStop}%
\bibitem [{\citenamefont {Gorshkov}\ \emph {et~al.}(2011)\citenamefont {Gorshkov}, \citenamefont {Manmana}, \citenamefont {Chen}, \citenamefont {Ye}, \citenamefont {Demler}, \citenamefont {Lukin},\ and\ \citenamefont {Rey}}]{Gorshkov2011a}%
  \BibitemOpen
  \bibfield  {author} {\bibinfo {author} {\bibfnamefont {A.~V.}\ \bibnamefont {Gorshkov}}, \bibinfo {author} {\bibfnamefont {S.~R.}\ \bibnamefont {Manmana}}, \bibinfo {author} {\bibfnamefont {G.}~\bibnamefont {Chen}}, \bibinfo {author} {\bibfnamefont {J.}~\bibnamefont {Ye}}, \bibinfo {author} {\bibfnamefont {E.}~\bibnamefont {Demler}}, \bibinfo {author} {\bibfnamefont {M.~D.}\ \bibnamefont {Lukin}},\ and\ \bibinfo {author} {\bibfnamefont {A.~M.}\ \bibnamefont {Rey}},\ }\bibfield  {title} {\bibinfo {title} {Tunable superfluidity and quantum magnetism with ultracold polar molecules},\ }\href {https://doi.org/10.1103/PhysRevLett.107.115301} {\bibfield  {journal} {\bibinfo  {journal} {Phys. Rev. Lett.}\ }\textbf {\bibinfo {volume} {107}},\ \bibinfo {pages} {115301} (\bibinfo {year} {2011})}\BibitemShut {NoStop}%
\bibitem [{\citenamefont {Yao}\ \emph {et~al.}(2018)\citenamefont {Yao}, \citenamefont {Zaletel}, \citenamefont {Stamper-Kurn},\ and\ \citenamefont {Vishwanath}}]{Yao2018}%
  \BibitemOpen
  \bibfield  {author} {\bibinfo {author} {\bibfnamefont {N.~Y.}\ \bibnamefont {Yao}}, \bibinfo {author} {\bibfnamefont {M.~P.}\ \bibnamefont {Zaletel}}, \bibinfo {author} {\bibfnamefont {D.~M.}\ \bibnamefont {Stamper-Kurn}},\ and\ \bibinfo {author} {\bibfnamefont {A.}~\bibnamefont {Vishwanath}},\ }\bibfield  {title} {\bibinfo {title} {A quantum dipolar spin liquid},\ }\href {https://doi.org/10.1038/s41567-017-0030-7} {\bibfield  {journal} {\bibinfo  {journal} {Nat.~Phys.}\ }\textbf {\bibinfo {volume} {14}},\ \bibinfo {pages} {405} (\bibinfo {year} {2018})}\BibitemShut {NoStop}%
\bibitem [{\citenamefont {Schmidt}\ \emph {et~al.}(2022)\citenamefont {Schmidt}, \citenamefont {Lassabli\`ere}, \citenamefont {Qu\'em\'ener},\ and\ \citenamefont {Langen}}]{Schmidt2022}%
  \BibitemOpen
  \bibfield  {author} {\bibinfo {author} {\bibfnamefont {M.}~\bibnamefont {Schmidt}}, \bibinfo {author} {\bibfnamefont {L.}~\bibnamefont {Lassabli\`ere}}, \bibinfo {author} {\bibfnamefont {G.}~\bibnamefont {Qu\'em\'ener}},\ and\ \bibinfo {author} {\bibfnamefont {T.}~\bibnamefont {Langen}},\ }\bibfield  {title} {\bibinfo {title} {Self-bound dipolar droplets and supersolids in molecular bose-einstein condensates},\ }\href {https://doi.org/10.1103/PhysRevResearch.4.013235} {\bibfield  {journal} {\bibinfo  {journal} {Phys. Rev. Res.}\ }\textbf {\bibinfo {volume} {4}},\ \bibinfo {pages} {013235} (\bibinfo {year} {2022})}\BibitemShut {NoStop}%
\bibitem [{\citenamefont {Albert}\ \emph {et~al.}(2020)\citenamefont {Albert}, \citenamefont {Covey},\ and\ \citenamefont {Preskill}}]{Albert_Preskill_2020_QEC}%
  \BibitemOpen
  \bibfield  {author} {\bibinfo {author} {\bibfnamefont {V.~V.}\ \bibnamefont {Albert}}, \bibinfo {author} {\bibfnamefont {J.~P.}\ \bibnamefont {Covey}},\ and\ \bibinfo {author} {\bibfnamefont {J.}~\bibnamefont {Preskill}},\ }\bibfield  {title} {\bibinfo {title} {Robust encoding of a qubit in a molecule},\ }\href {https://journals.aps.org/prx/abstract/10.1103/PhysRevX.10.031050} {\bibfield  {journal} {\bibinfo  {journal} {Phys. Rev. X}\ }\textbf {\bibinfo {volume} {10}},\ \bibinfo {pages} {031050} (\bibinfo {year} {2020})}\BibitemShut {NoStop}%
\bibitem [{\citenamefont {Sawant}\ \emph {et~al.}(2020)\citenamefont {Sawant}, \citenamefont {Blackmore}, \citenamefont {Gregory}, \citenamefont {Mur-Petit}, \citenamefont {Jaksch}, \citenamefont {Aldegunde}, \citenamefont {Hutson}, \citenamefont {Tarbutt},\ and\ \citenamefont {Cornish}}]{Sawant_Cornish_2020_Qudit}%
  \BibitemOpen
  \bibfield  {author} {\bibinfo {author} {\bibfnamefont {R.}~\bibnamefont {Sawant}}, \bibinfo {author} {\bibfnamefont {J.~A.}\ \bibnamefont {Blackmore}}, \bibinfo {author} {\bibfnamefont {P.~D.}\ \bibnamefont {Gregory}}, \bibinfo {author} {\bibfnamefont {J.}~\bibnamefont {Mur-Petit}}, \bibinfo {author} {\bibfnamefont {D.}~\bibnamefont {Jaksch}}, \bibinfo {author} {\bibfnamefont {J.}~\bibnamefont {Aldegunde}}, \bibinfo {author} {\bibfnamefont {J.~M.}\ \bibnamefont {Hutson}}, \bibinfo {author} {\bibfnamefont {M.~R.}\ \bibnamefont {Tarbutt}},\ and\ \bibinfo {author} {\bibfnamefont {S.~L.}\ \bibnamefont {Cornish}},\ }\bibfield  {title} {\bibinfo {title} {{Ultracold polar molecules as qudits}},\ }\href {https://doi.org/10.1088/1367-2630/ab60f4} {\bibfield  {journal} {\bibinfo  {journal} {New Journal of Physics}\ }\textbf {\bibinfo {volume} {22}},\ \bibinfo {pages} {013027} (\bibinfo {year} {2020})},\ \Eprint {https://arxiv.org/abs/1909.07484} {1909.07484} \BibitemShut {NoStop}%
\bibitem [{\citenamefont {Ni}\ \emph {et~al.}(2008)\citenamefont {Ni}, \citenamefont {Ospelkaus}, \citenamefont {de~Miranda}, \citenamefont {Pe'er}, \citenamefont {Neyenhuis}, \citenamefont {Zirbel}, \citenamefont {Kotochigova}, \citenamefont {Julienne}, \citenamefont {Jin},\ and\ \citenamefont {Ye}}]{Ni2008}%
  \BibitemOpen
  \bibfield  {author} {\bibinfo {author} {\bibfnamefont {K.-K.}\ \bibnamefont {Ni}}, \bibinfo {author} {\bibfnamefont {S.}~\bibnamefont {Ospelkaus}}, \bibinfo {author} {\bibfnamefont {M.~H.~G.}\ \bibnamefont {de~Miranda}}, \bibinfo {author} {\bibfnamefont {A.}~\bibnamefont {Pe'er}}, \bibinfo {author} {\bibfnamefont {B.}~\bibnamefont {Neyenhuis}}, \bibinfo {author} {\bibfnamefont {J.~J.}\ \bibnamefont {Zirbel}}, \bibinfo {author} {\bibfnamefont {S.}~\bibnamefont {Kotochigova}}, \bibinfo {author} {\bibfnamefont {P.~S.}\ \bibnamefont {Julienne}}, \bibinfo {author} {\bibfnamefont {D.~S.}\ \bibnamefont {Jin}},\ and\ \bibinfo {author} {\bibfnamefont {J.}~\bibnamefont {Ye}},\ }\bibfield  {title} {\bibinfo {title} {{A High Phase-Space-Density Gas of Polar Molecules}},\ }\href {https://doi.org/10.1126/science.1163861} {\bibfield  {journal} {\bibinfo  {journal} {Science}\ }\textbf {\bibinfo {volume} {322}},\ \bibinfo {pages} {231} (\bibinfo {year} {2008})}\BibitemShut {NoStop}%
\bibitem [{\citenamefont {Danzl}\ \emph {et~al.}(2008)\citenamefont {Danzl}, \citenamefont {Haller}, \citenamefont {Gustavsson}, \citenamefont {Mark}, \citenamefont {Hart}, \citenamefont {Bouloufa}, \citenamefont {Dulieu}, \citenamefont {Ritsch},\ and\ \citenamefont {N\"agerl}}]{Danzl2008}%
  \BibitemOpen
  \bibfield  {author} {\bibinfo {author} {\bibfnamefont {J.~G.}\ \bibnamefont {Danzl}}, \bibinfo {author} {\bibfnamefont {E.}~\bibnamefont {Haller}}, \bibinfo {author} {\bibfnamefont {M.}~\bibnamefont {Gustavsson}}, \bibinfo {author} {\bibfnamefont {M.~J.}\ \bibnamefont {Mark}}, \bibinfo {author} {\bibfnamefont {R.}~\bibnamefont {Hart}}, \bibinfo {author} {\bibfnamefont {N.}~\bibnamefont {Bouloufa}}, \bibinfo {author} {\bibfnamefont {O.}~\bibnamefont {Dulieu}}, \bibinfo {author} {\bibfnamefont {H.}~\bibnamefont {Ritsch}},\ and\ \bibinfo {author} {\bibfnamefont {H.-C.}\ \bibnamefont {N\"agerl}},\ }\bibfield  {title} {\bibinfo {title} {Quantum gas of deeply bound ground state molecules},\ }\href {https://doi.org/10.1126/science.1159909} {\bibfield  {journal} {\bibinfo  {journal} {Science}\ }\textbf {\bibinfo {volume} {321}},\ \bibinfo {pages} {1062} (\bibinfo {year} {2008})}\BibitemShut {NoStop}%
\bibitem [{\citenamefont {Lang}\ \emph {et~al.}(2008)\citenamefont {Lang}, \citenamefont {Winkler}, \citenamefont {Strauss}, \citenamefont {Grimm},\ and\ \citenamefont {Hecker~Denschlag}}]{Lang2008}%
  \BibitemOpen
  \bibfield  {author} {\bibinfo {author} {\bibfnamefont {F.}~\bibnamefont {Lang}}, \bibinfo {author} {\bibfnamefont {K.}~\bibnamefont {Winkler}}, \bibinfo {author} {\bibfnamefont {C.}~\bibnamefont {Strauss}}, \bibinfo {author} {\bibfnamefont {R.}~\bibnamefont {Grimm}},\ and\ \bibinfo {author} {\bibfnamefont {J.}~\bibnamefont {Hecker~Denschlag}},\ }\bibfield  {title} {\bibinfo {title} {Ultracold triplet molecules in the rovibrational ground state},\ }\href {https://doi.org/10.1103/PhysRevLett.101.133005} {\bibfield  {journal} {\bibinfo  {journal} {Phys. Rev. Lett.}\ }\textbf {\bibinfo {volume} {101}},\ \bibinfo {pages} {133005} (\bibinfo {year} {2008})}\BibitemShut {NoStop}%
\bibitem [{\citenamefont {Anderegg}\ \emph {et~al.}(2019)\citenamefont {Anderegg}, \citenamefont {Cheuk}, \citenamefont {Bao}, \citenamefont {Burchesky}, \citenamefont {Ketterle}, \citenamefont {Ni},\ and\ \citenamefont {Doyle}}]{Anderegg_Doyle_2019_Tweezer}%
  \BibitemOpen
  \bibfield  {author} {\bibinfo {author} {\bibfnamefont {L.}~\bibnamefont {Anderegg}}, \bibinfo {author} {\bibfnamefont {L.~W.}\ \bibnamefont {Cheuk}}, \bibinfo {author} {\bibfnamefont {Y.}~\bibnamefont {Bao}}, \bibinfo {author} {\bibfnamefont {S.}~\bibnamefont {Burchesky}}, \bibinfo {author} {\bibfnamefont {W.}~\bibnamefont {Ketterle}}, \bibinfo {author} {\bibfnamefont {K.-K.}\ \bibnamefont {Ni}},\ and\ \bibinfo {author} {\bibfnamefont {J.~M.}\ \bibnamefont {Doyle}},\ }\bibfield  {title} {\bibinfo {title} {{An optical tweezer array of ultracold molecules}},\ }\href {https://doi.org/10.1126/science.aax1265} {\bibfield  {journal} {\bibinfo  {journal} {Science}\ }\textbf {\bibinfo {volume} {365}},\ \bibinfo {pages} {1156} (\bibinfo {year} {2019})},\ \Eprint {https://arxiv.org/abs/1902.00497} {1902.00497} \BibitemShut {NoStop}%
\bibitem [{\citenamefont {Cairncross}\ \emph {et~al.}(2021)\citenamefont {Cairncross}, \citenamefont {Zhang}, \citenamefont {Picard}, \citenamefont {Yu}, \citenamefont {Wang},\ and\ \citenamefont {Ni}}]{cairncross_rovibgs_2021}%
  \BibitemOpen
  \bibfield  {author} {\bibinfo {author} {\bibfnamefont {W.~B.}\ \bibnamefont {Cairncross}}, \bibinfo {author} {\bibfnamefont {J.~T.}\ \bibnamefont {Zhang}}, \bibinfo {author} {\bibfnamefont {L.~R.~B.}\ \bibnamefont {Picard}}, \bibinfo {author} {\bibfnamefont {Y.}~\bibnamefont {Yu}}, \bibinfo {author} {\bibfnamefont {K.}~\bibnamefont {Wang}},\ and\ \bibinfo {author} {\bibfnamefont {K.-K.}\ \bibnamefont {Ni}},\ }\bibfield  {title} {\bibinfo {title} {Assembly of a rovibrational ground state molecule in an optical tweezer},\ }\href {https://doi.org/10.1103/PhysRevLett.126.123402} {\bibfield  {journal} {\bibinfo  {journal} {Physical Review Letters}\ }\textbf {\bibinfo {volume} {126}},\ \bibinfo {pages} {123402} (\bibinfo {year} {2021})}\BibitemShut {NoStop}%
\bibitem [{\citenamefont {Rosenberg}\ \emph {et~al.}(2022)\citenamefont {Rosenberg}, \citenamefont {Christakis}, \citenamefont {Guardado-Sanchez}, \citenamefont {Yan},\ and\ \citenamefont {Bakr}}]{rosenberg_observation_2022}%
  \BibitemOpen
  \bibfield  {author} {\bibinfo {author} {\bibfnamefont {J.~S.}\ \bibnamefont {Rosenberg}}, \bibinfo {author} {\bibfnamefont {L.}~\bibnamefont {Christakis}}, \bibinfo {author} {\bibfnamefont {E.}~\bibnamefont {Guardado-Sanchez}}, \bibinfo {author} {\bibfnamefont {Z.~Z.}\ \bibnamefont {Yan}},\ and\ \bibinfo {author} {\bibfnamefont {W.~S.}\ \bibnamefont {Bakr}},\ }\bibfield  {title} {\bibinfo {title} {Observation of the {Hanbury} {Brown}–{Twiss} effect with ultracold molecules},\ }\href {https://doi.org/10.1038/s41567-022-01695-9} {\bibfield  {journal} {\bibinfo  {journal} {Nature Physics}\ }\textbf {\bibinfo {volume} {18}},\ \bibinfo {pages} {1062} (\bibinfo {year} {2022})}\BibitemShut {NoStop}%
\bibitem [{\citenamefont {Ruttley}\ \emph {et~al.}(2023)\citenamefont {Ruttley}, \citenamefont {Guttridge}, \citenamefont {Spence}, \citenamefont {Bird}, \citenamefont {Le~Sueur}, \citenamefont {Hutson},\ and\ \citenamefont {Cornish}}]{Ruttley_Cornish_2023_Mergoassociation}%
  \BibitemOpen
  \bibfield  {author} {\bibinfo {author} {\bibfnamefont {D.~K.}\ \bibnamefont {Ruttley}}, \bibinfo {author} {\bibfnamefont {A.}~\bibnamefont {Guttridge}}, \bibinfo {author} {\bibfnamefont {S.}~\bibnamefont {Spence}}, \bibinfo {author} {\bibfnamefont {R.~C.}\ \bibnamefont {Bird}}, \bibinfo {author} {\bibfnamefont {C.~R.}\ \bibnamefont {Le~Sueur}}, \bibinfo {author} {\bibfnamefont {J.~M.}\ \bibnamefont {Hutson}},\ and\ \bibinfo {author} {\bibfnamefont {S.~L.}\ \bibnamefont {Cornish}},\ }\bibfield  {title} {\bibinfo {title} {Formation of ultracold molecules by merging optical tweezers},\ }\href {https://doi.org/10.1103/PhysRevLett.130.223401} {\bibfield  {journal} {\bibinfo  {journal} {Phys. Rev. Lett.}\ }\textbf {\bibinfo {volume} {130}},\ \bibinfo {pages} {223401} (\bibinfo {year} {2023})}\BibitemShut {NoStop}%
\bibitem [{\citenamefont {Liu}\ \emph {et~al.}(2018)\citenamefont {Liu}, \citenamefont {Hood}, \citenamefont {Yu}, \citenamefont {Zhang}, \citenamefont {Hutzler}, \citenamefont {Rosenband},\ and\ \citenamefont {Ni}}]{Liu2018}%
  \BibitemOpen
  \bibfield  {author} {\bibinfo {author} {\bibfnamefont {L.~R.}\ \bibnamefont {Liu}}, \bibinfo {author} {\bibfnamefont {J.~D.}\ \bibnamefont {Hood}}, \bibinfo {author} {\bibfnamefont {Y.}~\bibnamefont {Yu}}, \bibinfo {author} {\bibfnamefont {J.~T.}\ \bibnamefont {Zhang}}, \bibinfo {author} {\bibfnamefont {N.~R.}\ \bibnamefont {Hutzler}}, \bibinfo {author} {\bibfnamefont {T.}~\bibnamefont {Rosenband}},\ and\ \bibinfo {author} {\bibfnamefont {K.-K.}\ \bibnamefont {Ni}},\ }\bibfield  {title} {\bibinfo {title} {Building one molecule from a reservoir of two atoms},\ }\href {https://doi.org/10.1126/science.aar7797} {\bibfield  {journal} {\bibinfo  {journal} {Science}\ }\textbf {\bibinfo {volume} {360}},\ \bibinfo {pages} {900} (\bibinfo {year} {2018})}\BibitemShut {NoStop}%
\bibitem [{\citenamefont {Zhang}\ \emph {et~al.}(2022)\citenamefont {Zhang}, \citenamefont {Picard}, \citenamefont {Cairncross}, \citenamefont {Wang}, \citenamefont {Yu}, \citenamefont {Fang},\ and\ \citenamefont {Ni}}]{zhang_optical_2022}%
  \BibitemOpen
  \bibfield  {author} {\bibinfo {author} {\bibfnamefont {J.~T.}\ \bibnamefont {Zhang}}, \bibinfo {author} {\bibfnamefont {L.~R.~B.}\ \bibnamefont {Picard}}, \bibinfo {author} {\bibfnamefont {W.~B.}\ \bibnamefont {Cairncross}}, \bibinfo {author} {\bibfnamefont {K.}~\bibnamefont {Wang}}, \bibinfo {author} {\bibfnamefont {Y.}~\bibnamefont {Yu}}, \bibinfo {author} {\bibfnamefont {F.}~\bibnamefont {Fang}},\ and\ \bibinfo {author} {\bibfnamefont {K.-K.}\ \bibnamefont {Ni}},\ }\bibfield  {title} {\bibinfo {title} {An optical tweezer array of ground-state polar molecules},\ }\href {https://doi.org/10.1088/2058-9565/ac676c} {\bibfield  {journal} {\bibinfo  {journal} {Quantum Science and Technology}\ }\textbf {\bibinfo {volume} {7}},\ \bibinfo {pages} {035006} (\bibinfo {year} {2022})}\BibitemShut {NoStop}%
\bibitem [{\citenamefont {Picard}\ \emph {et~al.}(2023)\citenamefont {Picard}, \citenamefont {Zhang}, \citenamefont {Cairncross}, \citenamefont {Wang}, \citenamefont {Patenotte}, \citenamefont {Park}, \citenamefont {Yu}, \citenamefont {Liu}, \citenamefont {Hood}, \citenamefont {Gonz\'{a}lez-F\'{e}rez},\ and\ \citenamefont {Ni}}]{picard_high_2023}%
  \BibitemOpen
  \bibfield  {author} {\bibinfo {author} {\bibfnamefont {L.~R.~B.}\ \bibnamefont {Picard}}, \bibinfo {author} {\bibfnamefont {J.~T.}\ \bibnamefont {Zhang}}, \bibinfo {author} {\bibfnamefont {W.~B.}\ \bibnamefont {Cairncross}}, \bibinfo {author} {\bibfnamefont {K.}~\bibnamefont {Wang}}, \bibinfo {author} {\bibfnamefont {G.~E.}\ \bibnamefont {Patenotte}}, \bibinfo {author} {\bibfnamefont {A.~J.}\ \bibnamefont {Park}}, \bibinfo {author} {\bibfnamefont {Y.}~\bibnamefont {Yu}}, \bibinfo {author} {\bibfnamefont {L.~R.}\ \bibnamefont {Liu}}, \bibinfo {author} {\bibfnamefont {J.~D.}\ \bibnamefont {Hood}}, \bibinfo {author} {\bibfnamefont {R.}~\bibnamefont {Gonz\'{a}lez-F\'{e}rez}},\ and\ \bibinfo {author} {\bibfnamefont {K.-K.}\ \bibnamefont {Ni}},\ }\bibfield  {title} {\bibinfo {title} {{High resolution photoassociation spectroscopy of the excited $c^3\Sigma_{1}^+$ potential of $^{23}$Na$^{133}$Cs}},\ }\href {https://doi.org/10.1103/PhysRevResearch.5.023149} {\bibfield  {journal} {\bibinfo  {journal} {Phys. Rev.
  Research}\ }\textbf {\bibinfo {volume} {5}},\ \bibinfo {pages} {023149} (\bibinfo {year} {2023})}\BibitemShut {NoStop}%
\bibitem [{\citenamefont {Rosenband}\ \emph {et~al.}(2018)\citenamefont {Rosenband}, \citenamefont {Grimes},\ and\ \citenamefont {Ni}}]{Rosenband2018}%
  \BibitemOpen
  \bibfield  {author} {\bibinfo {author} {\bibfnamefont {T.}~\bibnamefont {Rosenband}}, \bibinfo {author} {\bibfnamefont {D.~D.}\ \bibnamefont {Grimes}},\ and\ \bibinfo {author} {\bibfnamefont {K.-K.}\ \bibnamefont {Ni}},\ }\bibfield  {title} {\bibinfo {title} {Elliptical polarization for molecular stark shift compensation in deep optical traps},\ }\href {https://doi.org/10.1364/OE.26.019821} {\bibfield  {journal} {\bibinfo  {journal} {Opt. Express}\ }\textbf {\bibinfo {volume} {26}},\ \bibinfo {pages} {19821} (\bibinfo {year} {2018})}\BibitemShut {NoStop}%
\bibitem [{\citenamefont {Picard}\ \emph {et~al.}(2024)\citenamefont {Picard}, \citenamefont {Patenotte}, \citenamefont {Park}, \citenamefont {Gebretsadkan},\ and\ \citenamefont {Ni}}]{picard_site-selective_2024}%
  \BibitemOpen
  \bibfield  {author} {\bibinfo {author} {\bibfnamefont {L.~R.~B.}\ \bibnamefont {Picard}}, \bibinfo {author} {\bibfnamefont {G.~E.}\ \bibnamefont {Patenotte}}, \bibinfo {author} {\bibfnamefont {A.~J.}\ \bibnamefont {Park}}, \bibinfo {author} {\bibfnamefont {S.~F.}\ \bibnamefont {Gebretsadkan}},\ and\ \bibinfo {author} {\bibfnamefont {K.-K.}\ \bibnamefont {Ni}},\ }\bibfield  {title} {\bibinfo {title} {Site-{Selective} {Preparation} and {Multistate} {Readout} of {Molecules} in {Optical} {Tweezers}},\ }\href {https://doi.org/10.1103/PRXQuantum.5.020344} {\bibfield  {journal} {\bibinfo  {journal} {PRX Quantum}\ }\textbf {\bibinfo {volume} {5}},\ \bibinfo {pages} {020344} (\bibinfo {year} {2024})}\BibitemShut {NoStop}%
\bibitem [{\citenamefont {Aymar}\ and\ \citenamefont {Dulieu}(2005)}]{Aymar2005}%
  \BibitemOpen
  \bibfield  {author} {\bibinfo {author} {\bibfnamefont {M.}~\bibnamefont {Aymar}}\ and\ \bibinfo {author} {\bibfnamefont {O.}~\bibnamefont {Dulieu}},\ }\bibfield  {title} {\bibinfo {title} {Calculation of accurate permanent dipole moments of the lowest {$^{1,3}\Sigma^+$} states of heteronuclear alkali dimers using extended basis sets},\ }\href {https://doi.org/10.1063/1.1903944} {\bibfield  {journal} {\bibinfo  {journal} {J. Chem. Phys.}\ }\textbf {\bibinfo {volume} {122}},\ \bibinfo {eid} {204302} (\bibinfo {year} {2005})}\BibitemShut {NoStop}%
\bibitem [{\citenamefont {Yan}\ \emph {et~al.}(2013)\citenamefont {Yan}, \citenamefont {Moses}, \citenamefont {Gadway}, \citenamefont {Covey}, \citenamefont {Hazzard}, \citenamefont {Rey}, \citenamefont {Jin},\ and\ \citenamefont {Ye}}]{Yan2013}%
  \BibitemOpen
  \bibfield  {author} {\bibinfo {author} {\bibfnamefont {B.}~\bibnamefont {Yan}}, \bibinfo {author} {\bibfnamefont {S.~A.}\ \bibnamefont {Moses}}, \bibinfo {author} {\bibfnamefont {B.}~\bibnamefont {Gadway}}, \bibinfo {author} {\bibfnamefont {J.~P.}\ \bibnamefont {Covey}}, \bibinfo {author} {\bibfnamefont {K.~R.~A.}\ \bibnamefont {Hazzard}}, \bibinfo {author} {\bibfnamefont {A.~M.}\ \bibnamefont {Rey}}, \bibinfo {author} {\bibfnamefont {D.~S.}\ \bibnamefont {Jin}},\ and\ \bibinfo {author} {\bibfnamefont {J.}~\bibnamefont {Ye}},\ }\bibfield  {title} {\bibinfo {title} {Observation of dipolar spin-exchange interactions with lattice-confined polar molecules},\ }\href {http://dx.doi.org/10.1038/nature12483} {\bibfield  {journal} {\bibinfo  {journal} {Nature}\ }\textbf {\bibinfo {volume} {501}},\ \bibinfo {pages} {521} (\bibinfo {year} {2013})}\BibitemShut {NoStop}%
\bibitem [{\citenamefont {de~Léséleuc}\ \emph {et~al.}(2017)\citenamefont {de~Léséleuc}, \citenamefont {Barredo}, \citenamefont {Lienhard}, \citenamefont {Browaeys},\ and\ \citenamefont {Lahaye}}]{deLeseleucPRL2017}%
  \BibitemOpen
  \bibfield  {author} {\bibinfo {author} {\bibfnamefont {S.}~\bibnamefont {de~Léséleuc}}, \bibinfo {author} {\bibfnamefont {D.}~\bibnamefont {Barredo}}, \bibinfo {author} {\bibfnamefont {V.}~\bibnamefont {Lienhard}}, \bibinfo {author} {\bibfnamefont {A.}~\bibnamefont {Browaeys}},\ and\ \bibinfo {author} {\bibfnamefont {T.}~\bibnamefont {Lahaye}},\ }\bibfield  {title} {\bibinfo {title} {Optical {Control} of the {Resonant} {Dipole}-{Dipole} {Interaction} between {Rydberg} {Atoms}},\ }\href {https://doi.org/10.1103/PhysRevLett.119.053202} {\bibfield  {journal} {\bibinfo  {journal} {Physical Review Letters}\ }\textbf {\bibinfo {volume} {119}},\ \bibinfo {pages} {053202} (\bibinfo {year} {2017})}\BibitemShut {NoStop}%
\bibitem [{\citenamefont {Wall}\ \emph {et~al.}(2015)\citenamefont {Wall}, \citenamefont {Hazzard},\ and\ \citenamefont {Rey}}]{Wall2015}%
  \BibitemOpen
  \bibfield  {author} {\bibinfo {author} {\bibfnamefont {M.~L.}\ \bibnamefont {Wall}}, \bibinfo {author} {\bibfnamefont {K.~R.~A.}\ \bibnamefont {Hazzard}},\ and\ \bibinfo {author} {\bibfnamefont {A.~M.}\ \bibnamefont {Rey}},\ }\bibinfo {title} {From atomic to mesoscale: The role of quantum coherence in systems of various complexities}\ (\bibinfo  {publisher} {World Scientific},\ \bibinfo {year} {2015})\ Chap.\ \bibinfo {chapter} {Quantum magnetism with ultracold molecules}\BibitemShut {NoStop}%
\bibitem [{\citenamefont {La~Porta}\ and\ \citenamefont {Wang}(2004)}]{la_porta_optical_2004}%
  \BibitemOpen
  \bibfield  {author} {\bibinfo {author} {\bibfnamefont {A.}~\bibnamefont {La~Porta}}\ and\ \bibinfo {author} {\bibfnamefont {M.~D.}\ \bibnamefont {Wang}},\ }\bibfield  {title} {\bibinfo {title} {Optical {Torque} {Wrench}: {Angular} {Trapping}, {Rotation}, and {Torque} {Detection} of {Quartz} {Microparticles}},\ }\href {https://doi.org/10.1103/PhysRevLett.92.190801} {\bibfield  {journal} {\bibinfo  {journal} {Physical Review Letters}\ }\textbf {\bibinfo {volume} {92}},\ \bibinfo {pages} {190801} (\bibinfo {year} {2004})}\BibitemShut {NoStop}%
\bibitem [{\citenamefont {Liu}\ \emph {et~al.}(2020)\citenamefont {Liu}, \citenamefont {Dong}, \citenamefont {Yang}, \citenamefont {Gong},\ and\ \citenamefont {Shi}}]{liu_robust_2020}%
  \BibitemOpen
  \bibfield  {author} {\bibinfo {author} {\bibfnamefont {W.}~\bibnamefont {Liu}}, \bibinfo {author} {\bibfnamefont {D.}~\bibnamefont {Dong}}, \bibinfo {author} {\bibfnamefont {H.}~\bibnamefont {Yang}}, \bibinfo {author} {\bibfnamefont {Q.}~\bibnamefont {Gong}},\ and\ \bibinfo {author} {\bibfnamefont {K.}~\bibnamefont {Shi}},\ }\bibfield  {title} {\bibinfo {title} {Robust and high‐speed rotation control in optical tweezers by using polarization synthesis based on heterodyne interference},\ }\href {https://doi.org/10.29026/oea.2020.200022} {\bibfield  {journal} {\bibinfo  {journal} {Opto-Electronic Advances}\ }\textbf {\bibinfo {volume} {3}},\ \bibinfo {pages} {200022} (\bibinfo {year} {2020})}\BibitemShut {NoStop}%
\bibitem [{\citenamefont {Souza}\ \emph {et~al.}(2012)\citenamefont {Souza}, \citenamefont {Álvarez},\ and\ \citenamefont {Suter}}]{souza_robust_2012}%
  \BibitemOpen
  \bibfield  {author} {\bibinfo {author} {\bibfnamefont {A.~M.}\ \bibnamefont {Souza}}, \bibinfo {author} {\bibfnamefont {G.~A.}\ \bibnamefont {Álvarez}},\ and\ \bibinfo {author} {\bibfnamefont {D.}~\bibnamefont {Suter}},\ }\bibfield  {title} {\bibinfo {title} {Robust dynamical decoupling},\ }\href {https://doi.org/10.1098/rsta.2011.0355} {\bibfield  {journal} {\bibinfo  {journal} {Philosophical Transactions of the Royal Society A: Mathematical, Physical and Engineering Sciences}\ }\textbf {\bibinfo {volume} {370}},\ \bibinfo {pages} {4748} (\bibinfo {year} {2012})}\BibitemShut {NoStop}%
\bibitem [{\citenamefont {Chomaz}\ \emph {et~al.}(2023)\citenamefont {Chomaz}, \citenamefont {Ferrier-Barbut}, \citenamefont {Ferlaino}, \citenamefont {Laburthe-Tolra}, \citenamefont {Lev},\ and\ \citenamefont {Pfau}}]{chomaz_dipolar_2023}%
  \BibitemOpen
  \bibfield  {author} {\bibinfo {author} {\bibfnamefont {L.}~\bibnamefont {Chomaz}}, \bibinfo {author} {\bibfnamefont {I.}~\bibnamefont {Ferrier-Barbut}}, \bibinfo {author} {\bibfnamefont {F.}~\bibnamefont {Ferlaino}}, \bibinfo {author} {\bibfnamefont {B.}~\bibnamefont {Laburthe-Tolra}}, \bibinfo {author} {\bibfnamefont {B.~L.}\ \bibnamefont {Lev}},\ and\ \bibinfo {author} {\bibfnamefont {T.}~\bibnamefont {Pfau}},\ }\bibfield  {title} {\bibinfo {title} {Dipolar physics: a review of experiments with magnetic quantum gases},\ }\href {https://doi.org/10.1088/1361-6633/aca814} {\bibfield  {journal} {\bibinfo  {journal} {Reports on Progress in Physics}\ }\textbf {\bibinfo {volume} {86}},\ \bibinfo {pages} {026401} (\bibinfo {year} {2023})}\BibitemShut {NoStop}%
\bibitem [{\citenamefont {Sackett}\ \emph {et~al.}(2000)\citenamefont {Sackett}, \citenamefont {Kielpinski}, \citenamefont {King}, \citenamefont {Langer}, \citenamefont {Meyer}, \citenamefont {Myatt}, \citenamefont {Rowe}, \citenamefont {Turchette}, \citenamefont {Itano}, \citenamefont {Wineland},\ and\ \citenamefont {Monroe}}]{sackett_experimental_2000}%
  \BibitemOpen
  \bibfield  {author} {\bibinfo {author} {\bibfnamefont {C.~A.}\ \bibnamefont {Sackett}}, \bibinfo {author} {\bibfnamefont {D.}~\bibnamefont {Kielpinski}}, \bibinfo {author} {\bibfnamefont {B.~E.}\ \bibnamefont {King}}, \bibinfo {author} {\bibfnamefont {C.}~\bibnamefont {Langer}}, \bibinfo {author} {\bibfnamefont {V.}~\bibnamefont {Meyer}}, \bibinfo {author} {\bibfnamefont {C.~J.}\ \bibnamefont {Myatt}}, \bibinfo {author} {\bibfnamefont {M.}~\bibnamefont {Rowe}}, \bibinfo {author} {\bibfnamefont {Q.~A.}\ \bibnamefont {Turchette}}, \bibinfo {author} {\bibfnamefont {W.~M.}\ \bibnamefont {Itano}}, \bibinfo {author} {\bibfnamefont {D.~J.}\ \bibnamefont {Wineland}},\ and\ \bibinfo {author} {\bibfnamefont {C.}~\bibnamefont {Monroe}},\ }\bibfield  {title} {\bibinfo {title} {Experimental entanglement of four particles},\ }\href {https://doi.org/10.1038/35005011} {\bibfield  {journal} {\bibinfo  {journal} {Nature}\ }\textbf {\bibinfo {volume} {404}},\ \bibinfo {pages} {256} (\bibinfo {year} {2000})}\BibitemShut
  {NoStop}%
\bibitem [{\citenamefont {Vexiau}\ \emph {et~al.}(2017)\citenamefont {Vexiau}, \citenamefont {Borsalino}, \citenamefont {Lepers}, \citenamefont {Orban}, \citenamefont {Aymar}, \citenamefont {Dulieu},\ and\ \citenamefont {Bouloufa-Maafa}}]{Vexiau2017}%
  \BibitemOpen
  \bibfield  {author} {\bibinfo {author} {\bibfnamefont {R.}~\bibnamefont {Vexiau}}, \bibinfo {author} {\bibfnamefont {D.}~\bibnamefont {Borsalino}}, \bibinfo {author} {\bibfnamefont {M.}~\bibnamefont {Lepers}}, \bibinfo {author} {\bibfnamefont {A.}~\bibnamefont {Orban}}, \bibinfo {author} {\bibfnamefont {M.}~\bibnamefont {Aymar}}, \bibinfo {author} {\bibfnamefont {O.}~\bibnamefont {Dulieu}},\ and\ \bibinfo {author} {\bibfnamefont {N.}~\bibnamefont {Bouloufa-Maafa}},\ }\bibfield  {title} {\bibinfo {title} {Dynamic dipole polarizabilities of heteronuclear alkali dimers: optical response, trapping and control of ultracold molecules},\ }\href {https://doi.org/10.1080/0144235X.2017.1351821} {\bibfield  {journal} {\bibinfo  {journal} {Int.~Rev.~Phys.~Chem.}\ }\textbf {\bibinfo {volume} {36}},\ \bibinfo {pages} {709} (\bibinfo {year} {2017})}\BibitemShut {NoStop}%
\bibitem [{\citenamefont {Gilmore}\ \emph {et~al.}(2021)\citenamefont {Gilmore}, \citenamefont {Affolter}, \citenamefont {Lewis-Swan}, \citenamefont {Barberena}, \citenamefont {Jordan}, \citenamefont {Rey},\ and\ \citenamefont {Bollinger}}]{gilmore2021quantum}%
  \BibitemOpen
  \bibfield  {author} {\bibinfo {author} {\bibfnamefont {K.~A.}\ \bibnamefont {Gilmore}}, \bibinfo {author} {\bibfnamefont {M.}~\bibnamefont {Affolter}}, \bibinfo {author} {\bibfnamefont {R.~J.}\ \bibnamefont {Lewis-Swan}}, \bibinfo {author} {\bibfnamefont {D.}~\bibnamefont {Barberena}}, \bibinfo {author} {\bibfnamefont {E.}~\bibnamefont {Jordan}}, \bibinfo {author} {\bibfnamefont {A.~M.}\ \bibnamefont {Rey}},\ and\ \bibinfo {author} {\bibfnamefont {J.~J.}\ \bibnamefont {Bollinger}},\ }\bibfield  {title} {\bibinfo {title} {{Quantum-enhanced sensing of displacements and electric fields with two-dimensional trapped-ion crystals}},\ }\href {https://doi.org/10.1126/science.abi5226} {\bibfield  {journal} {\bibinfo  {journal} {Science}\ }\textbf {\bibinfo {volume} {373}},\ \bibinfo {pages} {673} (\bibinfo {year} {2021})}\BibitemShut {NoStop}%
\bibitem [{\citenamefont {Koller}\ \emph {et~al.}(2015)\citenamefont {Koller}, \citenamefont {Mundinger}, \citenamefont {Wall},\ and\ \citenamefont {Rey}}]{koller2015demagnetization}%
  \BibitemOpen
  \bibfield  {author} {\bibinfo {author} {\bibfnamefont {A.~P.}\ \bibnamefont {Koller}}, \bibinfo {author} {\bibfnamefont {J.}~\bibnamefont {Mundinger}}, \bibinfo {author} {\bibfnamefont {M.~L.}\ \bibnamefont {Wall}},\ and\ \bibinfo {author} {\bibfnamefont {A.~M.}\ \bibnamefont {Rey}},\ }\bibfield  {title} {\bibinfo {title} {{Demagnetization dynamics of noninteracting trapped fermions}},\ }\href {https://doi.org/10.1103/PhysRevA.92.033608} {\bibfield  {journal} {\bibinfo  {journal} {Phys. Rev. A}\ }\textbf {\bibinfo {volume} {92}},\ \bibinfo {pages} {033608} (\bibinfo {year} {2015})}\BibitemShut {NoStop}%
\bibitem [{\citenamefont {Efron}\ and\ \citenamefont {Tibshirani}(1994)}]{efron_introduction_1994}%
  \BibitemOpen
  \bibfield  {author} {\bibinfo {author} {\bibfnamefont {B.}~\bibnamefont {Efron}}\ and\ \bibinfo {author} {\bibfnamefont {R.~J.}\ \bibnamefont {Tibshirani}},\ }\href {https://doi.org/10.1201/9780429246593} {\emph {\bibinfo {title} {An {Introduction} to the {Bootstrap}}}}\ (\bibinfo  {publisher} {Chapman and Hall/CRC},\ \bibinfo {address} {New York},\ \bibinfo {year} {1994})\BibitemShut {NoStop}%
\bibitem [{\citenamefont {Zhang}\ \emph {et~al.}(2020)\citenamefont {Zhang}, \citenamefont {Yu}, \citenamefont {Cairncross}, \citenamefont {Wang}, \citenamefont {Picard}, \citenamefont {Hood}, \citenamefont {Lin}, \citenamefont {Hutson},\ and\ \citenamefont {Ni}}]{Zhang_Ni_2020_MagnetoassociationTweezer}%
  \BibitemOpen
  \bibfield  {author} {\bibinfo {author} {\bibfnamefont {J.~T.}\ \bibnamefont {Zhang}}, \bibinfo {author} {\bibfnamefont {Y.}~\bibnamefont {Yu}}, \bibinfo {author} {\bibfnamefont {W.~B.}\ \bibnamefont {Cairncross}}, \bibinfo {author} {\bibfnamefont {K.}~\bibnamefont {Wang}}, \bibinfo {author} {\bibfnamefont {L.~R.~B.}\ \bibnamefont {Picard}}, \bibinfo {author} {\bibfnamefont {J.~D.}\ \bibnamefont {Hood}}, \bibinfo {author} {\bibfnamefont {Y.-W.}\ \bibnamefont {Lin}}, \bibinfo {author} {\bibfnamefont {J.~M.}\ \bibnamefont {Hutson}},\ and\ \bibinfo {author} {\bibfnamefont {K.-K.}\ \bibnamefont {Ni}},\ }\bibfield  {title} {\bibinfo {title} {Forming a single molecule by magnetoassociation in an optical tweezer},\ }\href {https://doi.org/10.1103/PhysRevLett.124.253401} {\bibfield  {journal} {\bibinfo  {journal} {Phys. Rev. Lett.}\ }\textbf {\bibinfo {volume} {124}},\ \bibinfo {pages} {253401} (\bibinfo {year} {2020})}\BibitemShut {NoStop}%
\bibitem [{\citenamefont {Chew}\ \emph {et~al.}(2022)\citenamefont {Chew}, \citenamefont {Tomita}, \citenamefont {Mahesh}, \citenamefont {Sugawa}, \citenamefont {de~L{\'e}s{\'e}leuc},\ and\ \citenamefont {Ohmori}}]{Chew2022}%
  \BibitemOpen
  \bibfield  {author} {\bibinfo {author} {\bibfnamefont {Y.}~\bibnamefont {Chew}}, \bibinfo {author} {\bibfnamefont {T.}~\bibnamefont {Tomita}}, \bibinfo {author} {\bibfnamefont {T.~P.}\ \bibnamefont {Mahesh}}, \bibinfo {author} {\bibfnamefont {S.}~\bibnamefont {Sugawa}}, \bibinfo {author} {\bibfnamefont {S.}~\bibnamefont {de~L{\'e}s{\'e}leuc}},\ and\ \bibinfo {author} {\bibfnamefont {K.}~\bibnamefont {Ohmori}},\ }\bibfield  {title} {\bibinfo {title} {Ultrafast energy exchange between two single rydberg atoms on a nanosecond timescale},\ }\href {https://doi.org/10.1038/s41566-022-01047-2} {\bibfield  {journal} {\bibinfo  {journal} {Nature Photonics}\ }\textbf {\bibinfo {volume} {16}},\ \bibinfo {pages} {724} (\bibinfo {year} {2022})}\BibitemShut {NoStop}%
\bibitem [{\citenamefont {Aldegunde}\ and\ \citenamefont {Hutson}(2017)}]{Aldegunde2017}%
  \BibitemOpen
  \bibfield  {author} {\bibinfo {author} {\bibfnamefont {J.}~\bibnamefont {Aldegunde}}\ and\ \bibinfo {author} {\bibfnamefont {J.~M.}\ \bibnamefont {Hutson}},\ }\bibfield  {title} {\bibinfo {title} {Hyperfine structure of alkali-metal diatomic molecules},\ }\href {https://doi.org/10.1103/PhysRevA.96.042506} {\bibfield  {journal} {\bibinfo  {journal} {Phys. Rev. A}\ }\textbf {\bibinfo {volume} {96}},\ \bibinfo {pages} {042506} (\bibinfo {year} {2017})}\BibitemShut {NoStop}%
\bibitem [{\citenamefont {Hofmann}(2005)}]{Hofmann2005_fidelity}%
  \BibitemOpen
  \bibfield  {author} {\bibinfo {author} {\bibfnamefont {H.~F.}\ \bibnamefont {Hofmann}},\ }\bibfield  {title} {\bibinfo {title} {Complementary classical fidelities as an efficient criterion for the evaluation of experimentally realized quantum operations},\ }\href {https://doi.org/10.1103/PhysRevLett.94.160504} {\bibfield  {journal} {\bibinfo  {journal} {Phys. Rev. Lett.}\ }\textbf {\bibinfo {volume} {94}},\ \bibinfo {pages} {160504} (\bibinfo {year} {2005})}\BibitemShut {NoStop}%
\bibitem [{\citenamefont {Sundar}\ \emph {et~al.}(2018)\citenamefont {Sundar}, \citenamefont {Gadway},\ and\ \citenamefont {Hazzard}}]{Sundar2018}%
  \BibitemOpen
  \bibfield  {author} {\bibinfo {author} {\bibfnamefont {B.}~\bibnamefont {Sundar}}, \bibinfo {author} {\bibfnamefont {B.}~\bibnamefont {Gadway}},\ and\ \bibinfo {author} {\bibfnamefont {K.~R.~A.}\ \bibnamefont {Hazzard}},\ }\bibfield  {title} {\bibinfo {title} {Synthetic dimensions in ultracold polar molecules},\ }\href {https://doi.org/10.1038/s41598-018-21699-x} {\bibfield  {journal} {\bibinfo  {journal} {Scientific Reports}\ }\textbf {\bibinfo {volume} {8}},\ \bibinfo {pages} {3422} (\bibinfo {year} {2018})}\BibitemShut {NoStop}%
\bibitem [{\citenamefont {Homeier}\ \emph {et~al.}(2024)\citenamefont {Homeier}, \citenamefont {Harris}, \citenamefont {Blatz}, \citenamefont {Geier}, \citenamefont {Hollerith}, \citenamefont {Schollw\"ock}, \citenamefont {Grusdt},\ and\ \citenamefont {Bohrdt}}]{Homeier2024}%
  \BibitemOpen
  \bibfield  {author} {\bibinfo {author} {\bibfnamefont {L.}~\bibnamefont {Homeier}}, \bibinfo {author} {\bibfnamefont {T.~J.}\ \bibnamefont {Harris}}, \bibinfo {author} {\bibfnamefont {T.}~\bibnamefont {Blatz}}, \bibinfo {author} {\bibfnamefont {S.}~\bibnamefont {Geier}}, \bibinfo {author} {\bibfnamefont {S.}~\bibnamefont {Hollerith}}, \bibinfo {author} {\bibfnamefont {U.}~\bibnamefont {Schollw\"ock}}, \bibinfo {author} {\bibfnamefont {F.}~\bibnamefont {Grusdt}},\ and\ \bibinfo {author} {\bibfnamefont {A.}~\bibnamefont {Bohrdt}},\ }\bibfield  {title} {\bibinfo {title} {Antiferromagnetic bosonic t-j models and their quantum simulation in tweezer arrays},\ }\href {https://doi.org/10.1103/PhysRevLett.132.230401} {\bibfield  {journal} {\bibinfo  {journal} {Phys. Rev. Lett.}\ }\textbf {\bibinfo {volume} {132}},\ \bibinfo {pages} {230401} (\bibinfo {year} {2024})}\BibitemShut {NoStop}%
\bibitem [{\citenamefont {Kuznetsova}\ \emph {et~al.}(2016)\citenamefont {Kuznetsova}, \citenamefont {Rittenhouse}, \citenamefont {Sadeghpour},\ and\ \citenamefont {Yelin}}]{kuznetsova_rydberg-atom-mediated_2016}%
  \BibitemOpen
  \bibfield  {author} {\bibinfo {author} {\bibfnamefont {E.}~\bibnamefont {Kuznetsova}}, \bibinfo {author} {\bibfnamefont {S.~T.}\ \bibnamefont {Rittenhouse}}, \bibinfo {author} {\bibfnamefont {H.~R.}\ \bibnamefont {Sadeghpour}},\ and\ \bibinfo {author} {\bibfnamefont {S.~F.}\ \bibnamefont {Yelin}},\ }\bibfield  {title} {\bibinfo {title} {Rydberg-atom-mediated nondestructive readout of collective rotational states in polar-molecule arrays},\ }\href {https://doi.org/10.1103/PhysRevA.94.032325} {\bibfield  {journal} {\bibinfo  {journal} {Phys. Rev. A}\ }\textbf {\bibinfo {volume} {94}},\ \bibinfo {pages} {032325} (\bibinfo {year} {2016})}\BibitemShut {NoStop}%
\bibitem [{\citenamefont {Wang}\ \emph {et~al.}(2022)\citenamefont {Wang}, \citenamefont {Williams}, \citenamefont {Picard}, \citenamefont {Yao},\ and\ \citenamefont {Ni}}]{wang_enriching_2022}%
  \BibitemOpen
  \bibfield  {author} {\bibinfo {author} {\bibfnamefont {K.}~\bibnamefont {Wang}}, \bibinfo {author} {\bibfnamefont {C.~P.}\ \bibnamefont {Williams}}, \bibinfo {author} {\bibfnamefont {L.~R.}\ \bibnamefont {Picard}}, \bibinfo {author} {\bibfnamefont {N.~Y.}\ \bibnamefont {Yao}},\ and\ \bibinfo {author} {\bibfnamefont {K.-K.}\ \bibnamefont {Ni}},\ }\bibfield  {title} {\bibinfo {title} {Enriching the {Quantum} {Toolbox} of {Ultracold} {Molecules} with {Rydberg} {Atoms}},\ }\href {https://doi.org/10.1103/PRXQuantum.3.030339} {\bibfield  {journal} {\bibinfo  {journal} {PRX Quantum}\ }\textbf {\bibinfo {volume} {3}},\ \bibinfo {pages} {030339} (\bibinfo {year} {2022})}\BibitemShut {NoStop}%
\bibitem [{\citenamefont {Zhang}\ and\ \citenamefont {Tarbutt}(2022)}]{Zhang_Tarbutt_2022_HybridMolRydb}%
  \BibitemOpen
  \bibfield  {author} {\bibinfo {author} {\bibfnamefont {C.}~\bibnamefont {Zhang}}\ and\ \bibinfo {author} {\bibfnamefont {M.}~\bibnamefont {Tarbutt}},\ }\bibfield  {title} {\bibinfo {title} {Quantum computation in a hybrid array of molecules and rydberg atoms},\ }\href {https://doi.org/10.1103/PRXQuantum.3.030340} {\bibfield  {journal} {\bibinfo  {journal} {PRX Quantum}\ }\textbf {\bibinfo {volume} {3}},\ \bibinfo {pages} {030340} (\bibinfo {year} {2022})}\BibitemShut {NoStop}%
\bibitem [{\citenamefont {Guttridge}\ \emph {et~al.}(2023)\citenamefont {Guttridge}, \citenamefont {Ruttley}, \citenamefont {Baldock}, \citenamefont {Gonz\'alez-F\'erez}, \citenamefont {Sadeghpour}, \citenamefont {Adams},\ and\ \citenamefont {Cornish}}]{Guttridge2023}%
  \BibitemOpen
  \bibfield  {author} {\bibinfo {author} {\bibfnamefont {A.}~\bibnamefont {Guttridge}}, \bibinfo {author} {\bibfnamefont {D.~K.}\ \bibnamefont {Ruttley}}, \bibinfo {author} {\bibfnamefont {A.~C.}\ \bibnamefont {Baldock}}, \bibinfo {author} {\bibfnamefont {R.}~\bibnamefont {Gonz\'alez-F\'erez}}, \bibinfo {author} {\bibfnamefont {H.~R.}\ \bibnamefont {Sadeghpour}}, \bibinfo {author} {\bibfnamefont {C.~S.}\ \bibnamefont {Adams}},\ and\ \bibinfo {author} {\bibfnamefont {S.~L.}\ \bibnamefont {Cornish}},\ }\bibfield  {title} {\bibinfo {title} {Observation of rydberg blockade due to the charge-dipole interaction between an atom and a polar molecule},\ }\href {https://doi.org/10.1103/PhysRevLett.131.013401} {\bibfield  {journal} {\bibinfo  {journal} {Phys. Rev. Lett.}\ }\textbf {\bibinfo {volume} {131}},\ \bibinfo {pages} {013401} (\bibinfo {year} {2023})}\BibitemShut {NoStop}%
\bibitem [{\citenamefont {Guardado-Sanchez}\ \emph {et~al.}(2021)\citenamefont {Guardado-Sanchez}, \citenamefont {Spar}, \citenamefont {Schauss}, \citenamefont {Belyansky}, \citenamefont {Young}, \citenamefont {Bienias}, \citenamefont {Gorshkov}, \citenamefont {Iadecola},\ and\ \citenamefont {Bakr}}]{guardado2021quench}%
  \BibitemOpen
  \bibfield  {author} {\bibinfo {author} {\bibfnamefont {E.}~\bibnamefont {Guardado-Sanchez}}, \bibinfo {author} {\bibfnamefont {B.~M.}\ \bibnamefont {Spar}}, \bibinfo {author} {\bibfnamefont {P.}~\bibnamefont {Schauss}}, \bibinfo {author} {\bibfnamefont {R.}~\bibnamefont {Belyansky}}, \bibinfo {author} {\bibfnamefont {J.~T.}\ \bibnamefont {Young}}, \bibinfo {author} {\bibfnamefont {P.}~\bibnamefont {Bienias}}, \bibinfo {author} {\bibfnamefont {A.~V.}\ \bibnamefont {Gorshkov}}, \bibinfo {author} {\bibfnamefont {T.}~\bibnamefont {Iadecola}},\ and\ \bibinfo {author} {\bibfnamefont {W.~S.}\ \bibnamefont {Bakr}},\ }\bibfield  {title} {\bibinfo {title} {{Quench Dynamics of a Fermi Gas with Strong Nonlocal Interactions}},\ }\href {https://doi.org/10.1103/PhysRevX.11.021036} {\bibfield  {journal} {\bibinfo  {journal} {Phys. Rev. X}\ }\textbf {\bibinfo {volume} {11}},\ \bibinfo {pages} {021036} (\bibinfo {year} {2021})}\BibitemShut {NoStop}%
\bibitem [{\citenamefont {Carroll}\ \emph {et~al.}(2024)\citenamefont {Carroll}, \citenamefont {Hirzler}, \citenamefont {Miller}, \citenamefont {Wellnitz}, \citenamefont {Muleady}, \citenamefont {Lin}, \citenamefont {Zamarski}, \citenamefont {Wang}, \citenamefont {Bohn}, \citenamefont {Rey},\ and\ \citenamefont {Ye}}]{carroll2024observation}%
  \BibitemOpen
  \bibfield  {author} {\bibinfo {author} {\bibfnamefont {A.~N.}\ \bibnamefont {Carroll}}, \bibinfo {author} {\bibfnamefont {H.}~\bibnamefont {Hirzler}}, \bibinfo {author} {\bibfnamefont {C.}~\bibnamefont {Miller}}, \bibinfo {author} {\bibfnamefont {D.}~\bibnamefont {Wellnitz}}, \bibinfo {author} {\bibfnamefont {S.~R.}\ \bibnamefont {Muleady}}, \bibinfo {author} {\bibfnamefont {J.}~\bibnamefont {Lin}}, \bibinfo {author} {\bibfnamefont {K.~P.}\ \bibnamefont {Zamarski}}, \bibinfo {author} {\bibfnamefont {R.~R.~W.}\ \bibnamefont {Wang}}, \bibinfo {author} {\bibfnamefont {J.~L.}\ \bibnamefont {Bohn}}, \bibinfo {author} {\bibfnamefont {A.~M.}\ \bibnamefont {Rey}},\ and\ \bibinfo {author} {\bibfnamefont {J.}~\bibnamefont {Ye}},\ }\href@noop {} {\bibinfo {title} {Observation of generalized t-j spin dynamics with tunable dipolar interactions}} (\bibinfo {year} {2024}),\ \Eprint {https://arxiv.org/abs/2404.18916} {arXiv:2404.18916} \BibitemShut {NoStop}%
\bibitem [{\citenamefont {Scholl}\ \emph {et~al.}(2023)\citenamefont {Scholl}, \citenamefont {Shaw}, \citenamefont {Finkelstein}, \citenamefont {Tsai}, \citenamefont {Choi},\ and\ \citenamefont {Endres}}]{scholl2023erasure_motion}%
  \BibitemOpen
  \bibfield  {author} {\bibinfo {author} {\bibfnamefont {P.}~\bibnamefont {Scholl}}, \bibinfo {author} {\bibfnamefont {A.~L.}\ \bibnamefont {Shaw}}, \bibinfo {author} {\bibfnamefont {R.}~\bibnamefont {Finkelstein}}, \bibinfo {author} {\bibfnamefont {R.~B.-S.}\ \bibnamefont {Tsai}}, \bibinfo {author} {\bibfnamefont {J.}~\bibnamefont {Choi}},\ and\ \bibinfo {author} {\bibfnamefont {M.}~\bibnamefont {Endres}},\ }\bibfield  {title} {\bibinfo {title} {{Erasure-cooling, control, and hyper-entanglement of motion in optical tweezers}},\ }\bibfield  {journal} {\bibinfo  {journal} {arXiv}\ }\href {https://doi.org/10.48550/arXiv.2311.15580} {10.48550/arXiv.2311.15580} (\bibinfo {year} {2023}),\ \Eprint {https://arxiv.org/abs/2311.15580} {2311.15580} \BibitemShut {NoStop}%
\bibitem [{\citenamefont {Boradjiev}\ and\ \citenamefont {Vitanov}(2013)}]{boradjiev_control_2013}%
  \BibitemOpen
  \bibfield  {author} {\bibinfo {author} {\bibfnamefont {I.~I.}\ \bibnamefont {Boradjiev}}\ and\ \bibinfo {author} {\bibfnamefont {N.~V.}\ \bibnamefont {Vitanov}},\ }\bibfield  {title} {\bibinfo {title} {Control of qubits by shaped pulses of finite duration},\ }\href {https://doi.org/10.1103/PhysRevA.88.013402} {\bibfield  {journal} {\bibinfo  {journal} {Physical Review A}\ }\textbf {\bibinfo {volume} {88}},\ \bibinfo {pages} {013402} (\bibinfo {year} {2013})}\BibitemShut {NoStop}%
\bibitem [{\citenamefont {Clopper}\ and\ \citenamefont {Pearson}(1934)}]{clopper_use_1934}%
  \BibitemOpen
  \bibfield  {author} {\bibinfo {author} {\bibfnamefont {C.~J.}\ \bibnamefont {Clopper}}\ and\ \bibinfo {author} {\bibfnamefont {E.~S.}\ \bibnamefont {Pearson}},\ }\bibfield  {title} {\bibinfo {title} {The use of confidence or fiducial limits illustrated in the case of the binomial},\ }\href {https://doi.org/10.1093/biomet/26.4.404} {\bibfield  {journal} {\bibinfo  {journal} {Biometrika}\ }\textbf {\bibinfo {volume} {26}},\ \bibinfo {pages} {404} (\bibinfo {year} {1934})}\BibitemShut {NoStop}%
\bibitem [{\citenamefont {Krämer}\ \emph {et~al.}(2018)\citenamefont {Krämer}, \citenamefont {Plankensteiner}, \citenamefont {Ostermann},\ and\ \citenamefont {Ritsch}}]{kramer_quantumopticsjl_2018}%
  \BibitemOpen
  \bibfield  {author} {\bibinfo {author} {\bibfnamefont {S.}~\bibnamefont {Krämer}}, \bibinfo {author} {\bibfnamefont {D.}~\bibnamefont {Plankensteiner}}, \bibinfo {author} {\bibfnamefont {L.}~\bibnamefont {Ostermann}},\ and\ \bibinfo {author} {\bibfnamefont {H.}~\bibnamefont {Ritsch}},\ }\bibfield  {title} {\bibinfo {title} {{QuantumOptics}.jl: {A} {Julia} framework for simulating open quantum systems},\ }\href {https://doi.org/10.1016/j.cpc.2018.02.004} {\bibfield  {journal} {\bibinfo  {journal} {Computer Physics Communications}\ }\textbf {\bibinfo {volume} {227}},\ \bibinfo {pages} {109} (\bibinfo {year} {2018})}\BibitemShut {NoStop}%
\bibitem [{\citenamefont {Singh}\ \emph {et~al.}(2009)\citenamefont {Singh}, \citenamefont {Senthilkumaran},\ and\ \citenamefont {Singh}}]{singh2009tight}%
  \BibitemOpen
  \bibfield  {author} {\bibinfo {author} {\bibfnamefont {R.~K.}\ \bibnamefont {Singh}}, \bibinfo {author} {\bibfnamefont {P.}~\bibnamefont {Senthilkumaran}},\ and\ \bibinfo {author} {\bibfnamefont {K.}~\bibnamefont {Singh}},\ }\bibfield  {title} {\bibinfo {title} {{Tight focusing of vortex beams in presence of primary astigmatism}},\ }\href {https://doi.org/10.1364/JOSAA.26.000576} {\bibfield  {journal} {\bibinfo  {journal} {J. Opt. Soc. Am. A, JOSAA}\ }\textbf {\bibinfo {volume} {26}},\ \bibinfo {pages} {576} (\bibinfo {year} {2009})}\BibitemShut {NoStop}%
\bibitem [{\citenamefont {Colbert}\ and\ \citenamefont {Miller}(1992)}]{colbert1992novel}%
  \BibitemOpen
  \bibfield  {author} {\bibinfo {author} {\bibfnamefont {D.~T.}\ \bibnamefont {Colbert}}\ and\ \bibinfo {author} {\bibfnamefont {W.~H.}\ \bibnamefont {Miller}},\ }\bibfield  {title} {\bibinfo {title} {{A novel discrete variable representation for quantum mechanical reactive scattering via the S{-}matrix Kohn method}},\ }\href {https://doi.org/10.1063/1.462100} {\bibfield  {journal} {\bibinfo  {journal} {J. Chem. Phys.}\ }\textbf {\bibinfo {volume} {96}},\ \bibinfo {pages} {1982} (\bibinfo {year} {1992})}\BibitemShut {NoStop}%
\bibitem [{\citenamefont {Holland}\ \emph {et~al.}(2024)\citenamefont {Holland}, \citenamefont {Lu}, \citenamefont {Li}, \citenamefont {Welsh},\ and\ \citenamefont {Cheuk}}]{holland2024demonstration}%
  \BibitemOpen
  \bibfield  {author} {\bibinfo {author} {\bibfnamefont {C.~M.}\ \bibnamefont {Holland}}, \bibinfo {author} {\bibfnamefont {Y.}~\bibnamefont {Lu}}, \bibinfo {author} {\bibfnamefont {S.~J.}\ \bibnamefont {Li}}, \bibinfo {author} {\bibfnamefont {C.~L.}\ \bibnamefont {Welsh}},\ and\ \bibinfo {author} {\bibfnamefont {L.~W.}\ \bibnamefont {Cheuk}},\ }\href@noop {} {\bibinfo {title} {Demonstration of erasure conversion in a molecular tweezer array}} (\bibinfo {year} {2024}),\ \Eprint {https://arxiv.org/abs/2406.02391} {arXiv:2406.02391} \BibitemShut {NoStop}%
\end{thebibliography}%

\appendix
\setcounter{figure}{0}
\renewcommand{\thefigure}{E.\arabic{figure}}

\newpage

\begin{table}
    \centering
    \begin{tabular}{c c c c}
         Transition  & Frequency - 2B (kHz) & Relative Rabi frequency (a.u.) \\ 
         \hline
         $\ket{0}\leftrightarrow\ket{3/2,5/2}\otimes\ket{1,0}$& 691.7(2)& 0.77(5)\\
         
$\ket{0}\leftrightarrow\ket{e}$& 0.2(1) & 1 \\

$\ket{0}\leftrightarrow\ket{3/2,5/2}\otimes\left(\frac{\ket{1,1}+\ket{1,-1}}{\sqrt{2}}\right)$& -710.3(2) & 0.84(2)\\

\hline 
$\ket{0}\leftrightarrow\ket{3/2,3/2}\otimes\left(\frac{\ket{1,1}-\ket{1,-1}}{\sqrt{2}}\right)$& 474.6(2) & 0.01838(7)\\
\hline 
$\ket{3/2,7/2}\otimes\ket{0,0}\leftrightarrow\ket{e}$& 466.1(3) & 0.0053(1)\\
$\ket{3/2,1/2}\otimes\ket{0,0}\leftrightarrow\ket{e}$& -927.1(3) & 0.0218(2)\\
$\ket{1}\leftrightarrow\ket{e}$& -1902.2(1) & 189(4)\\
\hline

    \end{tabular}
    \caption{Measured hyperfine transitions in NaCs. Eigenstates are labeled by their dominant nuclear spin and rotational quantum numbers, $\ket{m_{I_{Na}},m_{I_{Cs}}}\otimes\ket{N,m_N}$. The first three rows correspond to transitions from $\ket{0}$ to different rotational sublevels of $N=1$. The fourth row corresponds to a hyperfine-changing transition starting from $\ket{0}$, and the last three rows to hyperfine-changing transitions starting from $\ket{e}$. In all cases, the trap depth and magnetic field are fixed. Rabi frequency is defined relative to that of the $\ket{0}\leftrightarrow \ket{e}$ transition and corrected for the different microwave amplitudes used to probe each transition.}
    \label{tab:HyperfineTransitions}
\end{table}

\begin{figure}[h]
    \centering
    \includegraphics[width=1\columnwidth]{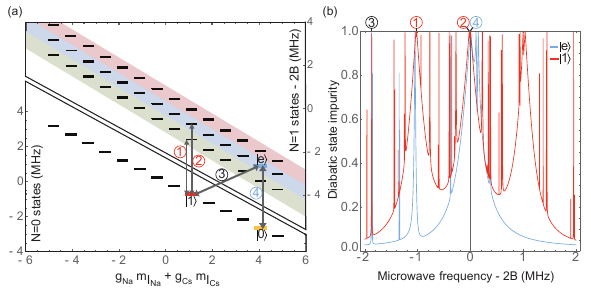}
    \caption{(a) Rotational-hyperfine structure of NaCs in the experimental regime. Color-shaded regions denote states of the same rotational sub-level. The x-axis represents the combined nuclear-spin contribution to the Zeeman shift, and is chosen to make the rotational structure apparent. (b) Estimated impurity of the states $\ket{1}$ and $\ket{e}$ when dressed by the microwave field used for the $\ket{1}\leftrightarrow\ket{e}$ transition, which depicts all available transitions from these states. The hyperfine-changing transition, labeled by 3, is red-detuned from $\ket{0}\leftrightarrow \ket{e}$, labelled 4, by 1.9 MHz and comparably narrow. Off-resonant coupling is primarily due to the transitions 1, 2, and 4.}
    \label{fig:hyperfineLevels}
\end{figure}

\begin{figure}[h]
    \centering
    \includegraphics[width=1\columnwidth]{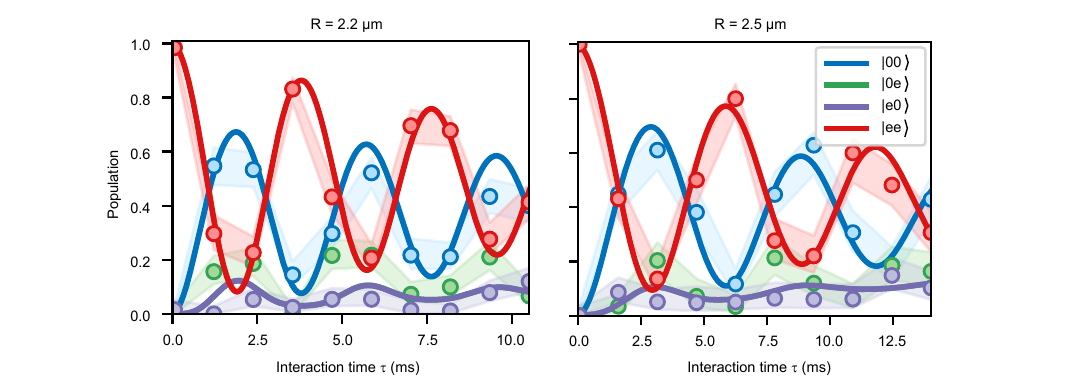}
    \caption{Dipole-dipole interactions using the same sequence as illustrated in Fig. \ref{fig:interactions} for distances of 2.2 and 2.5 \um.}
    \label{fig:distances}
\end{figure}

\begin{table}
    \centering
    \begin{tabular}{c c c c c c}
         Separation (\um)  & $J/h$ (Hz) & $\delta/h$ (Hz) & $\gamma_J$ (Hz) & $\gamma_\delta$ (Hz) & $\gamma_\Delta$ (Hz) \\ 
         \hline
         1.9  & 715(2) & 17(17) & 107(10)  & 25(3) & 13(2) \\ 
         2.2  & 477(3) & 214(111) & 125(29) & 31(31) & 16(16) \\ 
         2.5 & 314(6) & 159(159) & 114(32) & 53(53) & 35(35) \\ 
         \hline

    \end{tabular}
    \caption{Fit values and bootstrapped uncertainty estimates for master equation model of interactions, for the three distances shown in Fig. \ref{fig:interactions} and Extended Data Fig. \ref{fig:distances}. The bootstrapped distributions of fit values for the master equation dephasing rates and detuning $|\delta|$ are in some cases consistent with zero to within 1$\sigma$, and are truncated below zero due to the restriction that values must be positive. In these cases we report the value as 1$\sigma$ above zero, but note that the distribution deviates from Normal due to the truncation.}
    \label{tab:MasterEquationFits}
\end{table}

\begin{figure}[h]
    \centering
    \includegraphics[width=0.5\columnwidth]{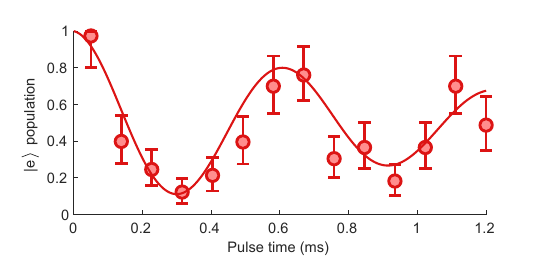}
    \caption{(a) Hyperfne-changing microwave pulse from $\ket{e}$ to $\ket{1}$. Fit is a $\cos^2$ function with symmetric exponential decay, giving a $\pi$-time of 0.307(9) ms and $1/e$ decay time of 1.2(4) ms.}
    \label{fig:hfpulse}
\end{figure}

\begin{figure}[h]
\label{fig:AppendixTruthtables}   
    \centering \includegraphics[width=\columnwidth]{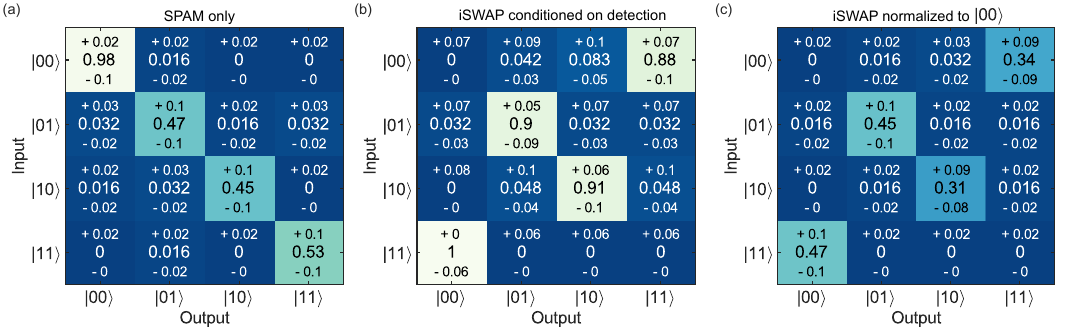}
    \caption{Full outcomes of truth table measurements for hyperfine gate: (a) Populations measured in each two-qubit state following state-preparation in that state, relative to population in $\ket{00}$, representing the relative SPAM fidelity of each state. (b) Populations measured in each two-qubit state as a fraction of the total detected molecule population in the $(\ket{0},\ket{1})$ manifold for each of the input states to the gate. This corrects for leakage to the $\ket{e}$ state during state-preparation and measurement, as well as during the hyperfine gate itself. (c) Populations measured in each two-qubit state following application of the iSWAP gate sequence, relative to the initial population in $\ket{00}$ in (a), illustrating the amount lost to leakage.}
    \label{fig:truthTables}
\end{figure}

\begin{figure}
    \centering
    \includegraphics[width=0.5\columnwidth]{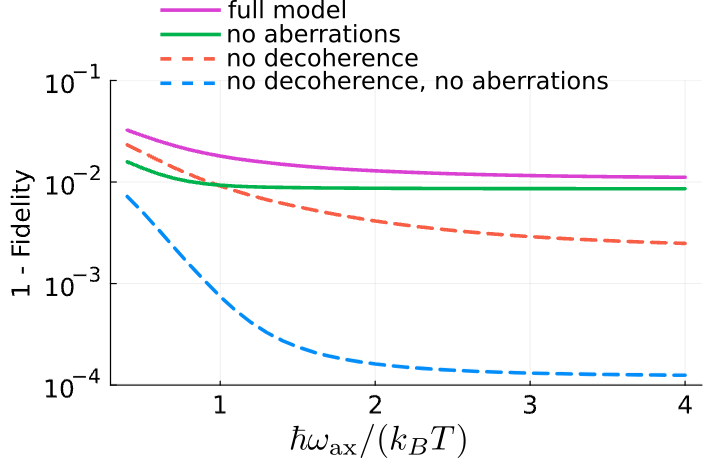}
    \caption{Bell state infidelity at the optimal time predicted as a function of inverse temperature.
    The solid purple curve illustrates the prediction of the full model including astigmatism of 0.16$\lambda$.
    The green curve illustrates the corresponding result without any astigmatism.
    The two dashed curves show what happens without single-molecule decoherence $\gamma_{\rm deph} = 0$, for both cases with (orange) and without (blue) astigmatism.
    Numbers quoted in the main text at 80\% ground state fraction correspond to $\hbar \omega_{\rm ax}/(k_B T) = 1.6$.
    Computed with $n_{\rm lvl} = 15$ ($n_{\rm lvl} = 10$ for $\hbar \omega_{\rm ax}/(k_B T) \geq 2$), leading to errors on the 0.1\% level at the lowest temperatures, and negligible errors for $\hbar \omega_{\rm ax}/(k_B T) > 1$.}
    \label{fig:bellFidelities}
\end{figure}

\end{document}